%% file: bcg_evolution.tex
\documentclass[numberedappendix]{emulateapj}
\usepackage{graphicx}
\usepackage{natbib}
\usepackage{apjfonts}
\usepackage{amsmath}
\usepackage{color}
\usepackage{comment}
\usepackage{rotating}

\bibpunct{(}{)}{;}{a}{}{,}

\begin{document}
\title{Galaxies in X-ray Selected Clusters and Groups in Dark Energy Survey Data I: \\ Stellar Mass Growth of Bright Central Galaxies  Since $z\sim1.2$}

\input{author_list.tex}

\email{\hspace{1em}$^\dagger$Email: Yuanyuan Zhang, ynzhang@umich.edu}

\begin{abstract}
Using the science verification data of the Dark Energy Survey (DES) for a new sample of 106 X-Ray selected clusters and groups, we study the stellar mass growth of Bright Central Galaxies (BCGs) since redshift 1.2. Compared with the expectation in a semi-analytical model applied to the Millennium Simulation, the observed BCGs become under-massive/under-luminous with decreasing redshift. We incorporate the uncertainties associated with cluster mass, redshift, and BCG stellar mass measurements into analysis of a redshift-dependent BCG-cluster mass relation,  $m_{*}\propto(\frac{M_{200}}{1.5\times 10^{14}M_{\odot}})^{0.24\pm 0.08}(1+z)^{-0.19\pm0.34}$, and compare the observed relation to the model prediction.  We estimate the average growth rate since $z = 1.0$ for BCGs hosted by clusters of $M_{200, z}=10^{13.8}M_{\odot}$, at $z=1.0$: $m_{*, BCG}$ appears to have grown by $0.13\pm0.11$ dex, in tension at $\sim 2.5 \sigma$ significance level with the $0.40$ dex growth rate expected from the semi-analytic model. We show that the buildup of extended intra-cluster light after $z=1.0$ may alleviate this tension in BCG growth rates. 
\end{abstract}
\maketitle

\section{Introduction}
Bright central galaxies (BCGs) are the luminous elliptical galaxies residing at the centers of galaxy clusters or groups. Once commonly referred to as the brightest cluster galaxies, the name {\it bright central galaxy} better reflects their special nature as the central galaxy of a massive halo. BCGs are surrounded by a subsidiary population of satellite galaxies. Their centrality and large size sets them apart from the general galaxy population. 

Early attention about BCGs started with studies about cD-type galaxies, since many BCGs are enveloped by extended stellar halos \citep{1964ApJ...140...35M}.
Statements that this population is not consistent with being statistically drawn from the global galaxy luminosity function led  \citet{1977ApJ...212..311T} to argue that BCGs require a special formation process.  Analytical and early numerical estimates of their  growth through dynamical friction and resultant cannibalism of cluster galaxies was soon identified as a viable process \citep{1975ApJ...202L.113O, 1976MNRAS.174...19W, 1978ApJ...224..320H,1983ApJ...268...30R}.  Early N-body simulations of merging pairs and groups of galaxies led \citet{1998ApJ...502..141D} to perform the first N-body study of BCG formation in a massive halo formed within a cold dark matter (CDM) cosmology. In that study, growth through early merging of a few massive galaxies dominated over late-time accretion of many smaller systems.  

The modern context of BCG assembly through hierarchical growth within an evolving spatial network of dark matter halos is now well established, but detailed understanding of various competing astrophysical processes remains elusive. Models in which BCGs accrete their stellar mass through ``dry'' merging with red and old galaxies produce scaling behavior and light profiles in fairly good agreement with observations \citep[{\it e.g.,} ][]{2009ApJ...696.1094R,2013MNRAS.435..901L}.   

Pure N-body models of dry merging ignore intra-cluster gas processes such as cooling and subsequent accretion and star formation of baryons onto the BCG.  Semi-analytical models find that such cooling needs to be mitigated by heating, and AGN feedback in a so-called ``radio mode'' is proposed as the solution \citep{2006MNRAS.367..864C, 2007MNRAS.375....2D}.  Simulations with explicit hydrodynamic treatment of the baryons are struggling to develop sub-grid models that capture the full complexity of the baryon behavior \cite[{\it e.g.,} ][]{2012MNRAS.420.2859M, 2013MNRAS.436.1750R, 2014MNRAS.443.1500M, 2014MNRAS.445.1774P}.

While BCG {\it in situ} star formation is almost certainly suppressed by the quenching effect of AGN (active galactic nuclei) feedback \citep{1994ARA&A..32..277F, 2012ARA&A..50..455F}, 
observational studies have found that residual star formation of $\sim 10-100\, M_{\odot}y\mathrm{r^{-1}}$ exists in many nearby BCGs \citep{2014MNRAS.444L..63F, 2012MNRAS.423..422L, 2014MNRAS.444..808G}.  A most puzzling study has observed a BCG starburst of $740\pm160\, M_{\odot}\mathrm{yr^{-1}}$ in the $z=0.596$ Phoenix cluster \citep{2012Natur.488..349M}.  Such a large star formation rate would contribute significantly to BCG stellar mass even if it lasted for just 1 Gyr.

Recent arguments based on local cooling--to--dynamical timescales  tie together this rich phenomenology in a self-regulated precipitation model \cite[][and references therein]{2015Natur.519..203V}.  Idealized hydrodynamic simulations  \citep{2014ApJ...789...54L, 2014ApJ...789..153L, 2015arXiv150302645M} support an episodic picture in which gas below a cooling threshold (roughly $t_{\rm cool} / t_{\rm dyn} < 10$) feeds black hole accretion and local star formation, with AGN feedback serving as the rectifier that shuts down cooling and allows the cycle to refresh. With {\it HST} observations of BCGs in the {\it CLASH} sample, \citet{2015arXiv150400598D} offer evidence that ultraviolet morphologies and star-formation rates of BCGs in CLASH clusters display features remarkably similar to those anticipated by these simulations.   

The semi-analytical expectations of BCG growth have been called into question by a number of observations that report significantly slower build-up of stellar mass over time \citep{2008MNRAS.387.1253W, 2009Natur.458..603C, 2012MNRAS.427..550L, 2013ApJ...771...61L, 2013ApJ...771...61L, 2014MNRAS.440..762O, 2015MNRAS.446.1107I}. This tension highlights limitations in our current understanding of BCG formation and motivates the work in this paper.

The production of intra-cluster light (ICL) is another important process affecting BCG formation over time. The ICL contains stars that got dispersed into intracluster space from BCGs or BCG mergers \citep[see:][]{2014MNRAS.437.3787C}. Simulation and observational studies show that ICL can make up 5-50\% of the total cluster/group stellar content \citep{2005MNRAS.358..949Z, 2006AJ....131..168K, 2007AJ....134..466K, 2007ApJ...666..147G, 2011MNRAS.414..602T, 2012A&A...537A..64G, 2012MNRAS.425.2058B, 2014ApJ...794..137M, 2014ApJ...781...24G, 2014A&A...565A.126P, 2015arXiv150304321B}. Details of how the ICL is formed and how its properties might vary from cluster to cluster remain unsettled  \citep{2006ApJ...652L..89M, 2007ApJ...668..826C, 2010MNRAS.406..936P, 2011ApJ...732...48R, 2014MNRAS.437..816C, 2014MNRAS.437.3787C, 2015arXiv150102251D}.

To advance our understanding about the above processes and BCG formation in general, it is important that we continue to refine our measurements of BCG growth. Most up-to-date observations are yielding perplexing or even contradictory results on this subject, perhaps because of in-comparability in their processing BCG observables \citep{2005MNRAS.361.1287M, 2007ApJ...662..808L, 2007AJ....133.1741B}. For instance, a few studies based on high redshift (z > 1.0) X-ray selected clusters \citep{2010ApJ...718...23S, 2011MNRAS.414..445S, 2009Natur.458..603C, 2008MNRAS.387.1253W} finds no sign of BCG stellar mass growth, while others based on clusters at low and high redshifts do observe the change \citep[which included some of the samples from the forementioned X-ray studies]{2002MNRAS.329L..53B, 2013ApJ...771...61L, 2012MNRAS.427..550L}. On the other hand, deriving BCG luminosity and hence BCG stellar mass from imaging data is not straightforward, and inconsistent measurements may have affected many previous findings about BCG formation. Finally, BCG mass is known to be correlated with cluster mass, which needs to accounted for when studying the change of BCG mass over time \citep[see for example,][]{2012MNRAS.427..550L}. Advances in our understanding of the nature of the growth of BCGs require a careful accounting of all of the ingredients, including their measurement uncertainties. 

In this paper, we investigate BCG stellar mass growth using DES Science Verification (DES SV) data, and a new sample of 106 X-ray selected clusters and groups from the DES {\it XMM} Cluster Survey (XCS), an {\it XMM-Newton} archival discovery project. Through using this X-ray selected sample, selection effect on studying BCG's optical properties are greatly alleviated: X-ray selected clusters display a wider variety of optical properties compared to optically selected clusters \citep{2012ApJ...752...12H}. The cluster and group sample spans a redshift range of $[0, 1.2]$, and a mass range of $[3\times10^{13}M_{\odot}, ~2\times10^{15}M_{\odot}]$. While most previous studies on this redshift range or cluster mass range are combining different samples or different imaging data sets, we study a single cluster sample with the deep optical data from DES. In this paper, we also pay particular attention to possible biases affecting BCG photometry, and have carefully evaluated the uncertainties associated with cluster mass, redshift, BCG luminosity and BCG stellar mass measurements. We provide details of our uncertainty, bias and covariance estimations in Appendices \ref{sec:mass}  to \ref{sec:additional_likelihood}. 

The rest of this paper is organized in the following order. In Section~\ref{sec:data}, we present our data sets and derive cluster masses, BCG luminosities and BCG stellar masses. We perform a matching exercise of BCG redshift evolution to the Millennium Simulation expectations in Section~\ref{sec:comparison}, then fit both simulated and observed BCG populations to a simple low-order model in Section~\ref{sec:shm}. We compare this model to previous estimates of BCG growth rate in Section~\ref{sec:rate}. We summarize our results in Section~\ref{sec:summary}.  Appendices \ref{sec:mass}  to \ref{sec:additional_likelihood} describe the uncertainties, biases and covariances of relevant measurements. Throughout this paper, we assume $\Omega_m$ to be 0.3, $\Omega_\Lambda$ to be 0.7, and the Hubble parameter $h$ to be 0.7.

\section{Data}
\label{sec:data}

This paper is based on an X-ray selected cluster and group sample from the DES-XCS project. BCG photometry is derived from DES Science Verification data. 

The rest of this section introduces the DES-XCS sample and the DES SV data, and also summarizes our procedures of deriving cluster masses, selecting BCGs and measuring BCG properties. Appendices  \ref{sec:mass}  to \ref{sec:bcg_lm} should be considered as extensions of this section.

\subsection{DES Science Verification Data}

The Dark Energy Survey is a ground-based optical survey that uses the wide-field DECam camera \citep{2015AJ....150..150F} mounted on the 4m Blanco telescope to image 5,000 $\mathrm{deg}^2$ of the southern hemisphere sky \citep{2010JPhCS.259a2080S}. The paper is based on 200 $\mathrm{deg}^2$ DES Science Verification (SV) data. This data set was taken during the 2012B observing season before the main survey \citep{diehl2014dark} began. A large fraction of the SV data have full DES imaging depth \citep{2013AAS...22135226L} and are processed with the official DES data processing pipeline  \citep{2012SPIE.8451E..0DM}. A more detailed review can be found in \cite{2014MNRAS.445.1482S}.

\subsection{The DES-XCS Cluster and Group Sample}
\begin{figure}
\includegraphics[width=0.5\textwidth]{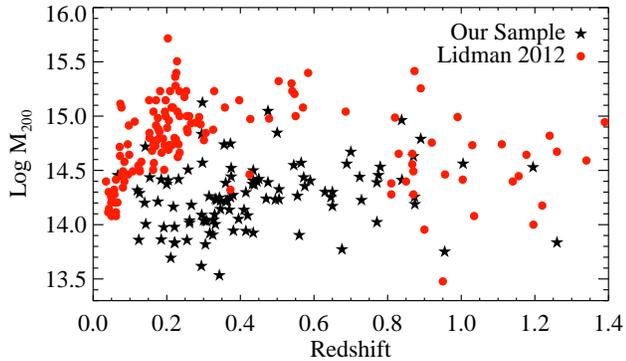}
\caption{The mass and redshift distribution of the DES-XCS sample (black stars) compared to that of Lidman et al. (2012, red circles).}
\label{fig:mass_z}
\end{figure}

The XMM Cluster Survey serendipitously searches for galaxy cluster (and group) candidates in the {\it XMM-Newton} archive \citep{2011MNRAS.418...14L, 2012MNRAS.423.1024M, 2013AN....334..462V}. The cluster candidates are then verified with optical/infrared imaging data, which confirm the existence of red sequence galaxies. Photometric redshifts of the confirmed clusters are also subsequently derived with the red sequence locus. Using DES SV data, \nocite{2014Miller} Miller et al. (in prep.; referred to as M15 in the rest of the paper) have identified $\sim170$ X-ray selected clusters and groups from XCS. M15  also measures their photometric redshifts and verify the measurements against archival spectroscopic redshifts \footnote{\texttt{http://ned.ipac.caltech.edu}}. In this paper, we use a sub-sample from  M15 that consists of 106 clusters and groups with mass above $3.0\times 10^{13}\mathrm{M_{\odot}}$. These clusters and groups are all referred to as  "clusters" in the rest of the paper. In Figure~\ref{fig:mass_z}, we show their mass and redshift distribution. For comparison, we also show the mass and redshift distribution of the cluster sample used in a similar study \citep{2012MNRAS.427..550L}. Our sample covers a lower mass range, and appears to be more evenly distributed in the redshift-mass space.

The cluster mass ($M_{200}$, the mass inside a 3D aperture within which the averaged matter density is 200 times the critical density) is either derived with X-ray temperature or X-ray luminosity, using a lensing calibrated $M-T$ relation \citep{2013ApJ...778...74K}. Because XCS is a serendipitous survey, not all the clusters have high quality X-ray temperature measurements. For these clusters, we derive their masses from X-ray luminosity. Further details about this procedure, and about the mass uncertainties can be found in Appendix~\ref{sec:mass}. 

We note that a handful of the clusters do not seem to have significant galaxy over-density associated with them. It is possible that our sample contains spurious clusters which originates from foreground/background X-ray contaminations. We have re-analyzed our analysis after removing 8 clusters that are not associated with significant galaxy over-density. The results are consistent with those presented in this paper within 0.5 $\sigma$. Given that these 8 clusters are in the low mass range (generally below $10^{14}~M_\odot$), removing them may introduce an artificial mass selection effect. We therefore do not attempt to do so in this paper. We also note that other factors, including BCG photometry measurement and cluster mass scaling relations at the low mass end (see discussion in Section~\ref{sec:constraint_shm} and \ref{sec:growth_result}), have bigger effect on our results than the possible spurious clusters in the sample.

\subsection{BCG Selection}
\begin{figure}
\includegraphics[width=0.5\textwidth]{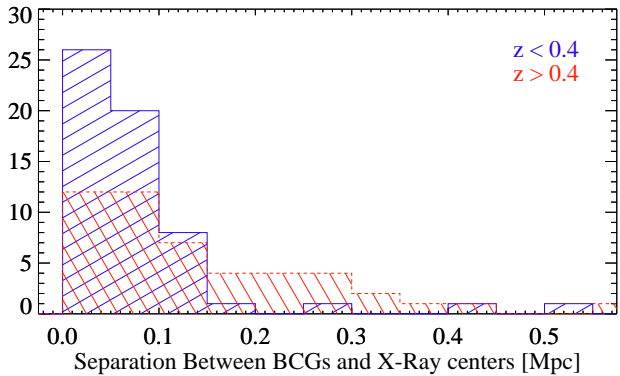}
\caption{ Distances between the BCGs and the X-ray emission centers for our cluster sample. Half of the BCGs are separated less than 0.07 Mpc (transverse comoving distance with negligible uncertainties from redshifts measurements) from the X-ray centers, and the large separations (> 0.4 Mpc) happen in clusters that may not be relaxed or appear to have spurious foreground/background emissions.}
\label{fig:dist_bcg}
\end{figure}

The BCGs are selected through visually examining the DES optical images, the X-ray emission contours, and the galaxy color-magnitude diagram. In this procedure, we aim to select a bright, extended, elliptical galaxy close to the X-ray emission center, which also roughly lies on the cluster red sequence. If there exist several red, equally bright and extended ellipticals close to the X-ray center, we select the nearest one. We did not notice a proper BCG candidate with a blue color.

We check our visual BCG selection against the central galaxy choice of a preliminary version of the DES SV RedMaPPer cluster catalog \citep[see the algorithm in][]{2014ApJ...785..104R}. Out of the 106 XCS clusters and groups, 64 are matched to RedMaPPer clusters and the majority (61) identify the same BCG. In the cases where we disagree with the BCG, we choose the brighter, more extended galaxy closest to the X-ray center while RedMaPPer selects a galaxy further away. The other 42 non-matches are caused by the different data coverage, redshift limit, and mass selection of the two catalogs: the RedMaPPer catalog employs only a subset of the SV data to achieve relatively uniform depth for selecting rich clusters below redshift 0.9. 

In Figure~\ref{fig:dist_bcg}, we show the distance distribution between the selected BCGs and the X-ray emission centers. Half of the BCGs are separated by less than 0.07 Mpc \citep[comparable to][]{2004ApJ...617..879L} from the X-ray emission centers, regardless of the redshifts of the clusters.

\subsection{BCG Photometry, Luminosity and Stellar Mass}
\label{sec:bcg_phot}

Measuring BCG photometry is among the {\it most} controversial topics in BCG studies. In Appendix~\ref{sec:phot}, we discuss complications and possible biases associated with Petrosian magnitude, Kron magnitude, profile fitting magnitude and aperture magnitude with extended details. We use magnitude measured with circular apertures of 15 kpc, 32 kpc, 50 kpc and 60 kpc radii. The main results are derived with the 32 kpc radius apertures, considering the BCG half light radius measurements in \cite{2011MNRAS.414..445S}. Detailed rationalization about this choice and description about our measurement procedure can also be found in Appendix~\ref{sec:phot}. 

We correct for galactic extinction using the stellar locus regression method \citep{2009AJ....138..110H, 2014MNRAS.439...28K, Rykoff}, and compute BCG luminosities and stellar masses using the stellar population modeling technique. We employ a 
\cite{2003PASP..115..763C} Initial Mass Function (IMF) and the \cite{2009ApJ...699..486C,  2010ApJ...712..833C} simple stellar population (SSP) models to construct stellar population templates, and select templates according to BCG DES $g,r,i,z$ photometry.
We use the best-fit model to compute the K-correction factor and the mass-to-light ratio. We evaluate uncertainties associated with BCG apparent magnitude, redshift, and BCG mass-to-light ratio. Further details about these procedures can be found in Appendix~\ref{sec:bcg_lm}.

\section{Simulation Matching Analysis}
\label{sec:comparison}

\begin{figure}
\includegraphics[width=0.5\textwidth]{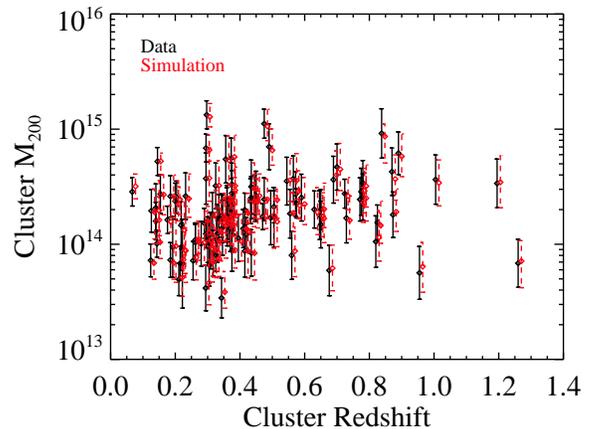}
\caption{ Mass and redshift distribution of the cluster sample used for this paper and distribution of the simulation clusters drawn from DL07. The black data points show the masses and mass uncertainties of the XCS clusters. The red data points show the median masses and the 0.158 and 0.842 percentiles of the simulation clusters. For clarification, we show the mass distribution of the resampled DL07 clusters at the redshift of the corresponding XCS cluster with a small offset. }
\label{fig:draw_sim}
\end{figure}

\begin{figure}
\includegraphics[width=0.5\textwidth]{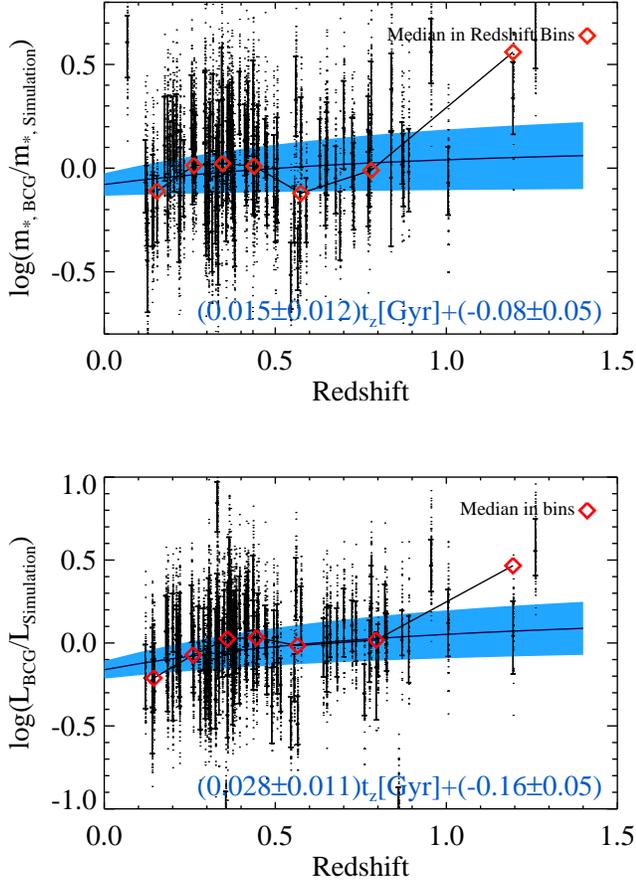}
\caption{ Comparison between the observed and simulated BCG properties. We show the median and 0.158, 0.842 percentiles of the differences. The dots are data points outside the range of the uncertainty whiskers, and the red diamonds are moving medians. The blue bands show the fitted linear model with dependence on look-back time, encompassing 1$\sigma$ uncertainties. We notice that the observed BCGs are becoming under-massive/under-luminous at decreasing redshift. Note that we have not included stellar mass or luminosity measurement uncertainties in the linear fit. We also use a 95\% confidence interval clipping technique to remove outliers.}
\label{fig:mass_sim}
\end{figure}

We first inspect the redshift evolution of BCG luminosity and stellar mass through matching our data with a semi-analytical simulation. We compare BCG luminosities and stellar masses to the corresponding values in the simulation, with diagrams analogous to those presented in many previous studies \citep{ 2009Natur.458..603C, 2014MNRAS.439.1294L, 2012MNRAS.427..550L, 2012ApJ...759...43T, 2013ApJ...771...61L, 2014MNRAS.440..762O} that overlay redshift evolutions of the observed and simulated BCG properties. The simulation involved in this comparison is the \cite{2007MNRAS.375....2D} semi-analytical (SAM) simulation (referred as DL07 hereafter) based on the Millennium project \citep{2005Natur.435..629S, 2013MNRAS.428.1351G}.  

\subsection{Simulation Sample Selection}
\label{sec:sim_select}

Since BCG luminosity and stellar mass are known to be correlated with cluster mass, the comparison between observation and simulation need to be made between clusters of similar masses. For each BCG in our sample, we compare it to a simulation subsample of 100 BCGs hosted by clusters of similar masses and redshifts. The simulation data are selected with the following procedure.
\begin{enumerate}
\item Identify simulation clusters with redshifts closest to that of the XCS cluster. Ideally,  we would have identified a cluster sub-sample with their redshift distribution matching the redshift uncertainty of the XCS cluster, but this is not possible since simulations are stored at discrete redshifts.

\item Select from the redshift sub-sample of 100 clusters with their posterior mass distribution (log-normal) matching the mass uncertainty of the XCS cluster. Note that we are not using the cluster mass function as a prior. Application of this prior leads to sampling clusters $\sim$ 0.1 dex less massive, but leave the conclusions unchanged.
\end{enumerate}

Note that in the above procedure, we are not considering additional cluster properties beyond $M_{200}$ and redshift. There is emerging evidence that X-ray selected clusters may be biased in terms of cluster concentration distribution \citep{2013ApJ...776...39R}, but it is un-clear how the bias would affect BCG formation study. We also do not consider the Eddington bias associated with $L_{X}$. The $M_{200}$ of the lowest $L_{X}$/$T_{X}$ systems are derived with $T_{X}$. Future studies yielding higher precision on BCG growth may wish to take these selection effects into consideration.

In Figure~\ref{fig:draw_sim}, we show the redshift and the mass distribution  of the XCS clusters together with the re-sampled DL07 simulation clusters. The above procedure produces a simulation sub-sample that well resembles the probability distribution of the XCS sample.

\subsection{Redshift Evolution of the Observed BCGs}
\label{sec:loz}

We directly compute the relative luminosity and stellar mass difference between the observed and simulated BCGs, as shown in Figure~\ref{fig:mass_sim}.  \footnote{We are comparing the observer frame DES {\it z} band luminosity to the observer frame SDSS {\it z} band luminosity in DL07. The response curves of the DES {\it z} band and the SDSS {\it z} band are similar enough, that the magnitude measurements for one object in the two systems shall be close within 0.05 mag.  We have tested this statement through cross matching galaxies in the SDSS stripe 82 database and the DES Year 1 coadd database. Although it is possible to transform between DES {\it z} band  magnitudes and SDSS {\it z} band magnitudes,  we avoid doing so because the transformation inevitably makes assumption about BCG SEDs. }

We notice that the differences between the observed and simulated BCGs change with redshift. The effect suggests that the observed BCGs do not grow as rapidly as in DL07 -- a different redshift evolution history in the observation. We fit the differences with a linear dependence on lookback time: if the redshift evolution of the observed BCGs is consistent with that in DL07, the slope of the linear fit shall be 0. This null hypothesis is not favored.

In Figure~\ref{fig:mass_sim}, we show the linear fitting result with blue bands which encompass the 1$\sigma$ uncertainties. The luminosity redshift evolution in the observation is different from the simulation with a 2.5 $\sigma$ significance ($0.028\pm0.011$). The significance from the stellar mass comparison is lower at 1.3 $\sigma$ ($0.015\pm0.012$), but BCG stellar mass is less certain (recall that it requires a choice for the mass-to-light ratio) and therefore the result is noisier.

The redshift evolution difference shows that the observed BCGs become increasingly under-massive/under-luminous at decreasing redshift  compared to DL07 \citep[compare the result to][]{2012MNRAS.427..550L, 2013ApJ...771...61L, 2014MNRAS.440..762O}. At the lowest redshift bin ($z \sim 0.1$) in Figure~\ref{fig:mass_sim}, the observed BCGs appear to be 0.1 to 0.2 dex \footnote{ $x~\mathrm{dex}=10^x$} under-massive/under-luminous as a result of a different redshift evolution history.

Arguably, the above statement relies on a fitting function connecting the difference between the observed and simulated BCG properties to redshift. The significance level of this statement depends on the exact form of the fitting function. In Section ~\ref{sec:shm} and~\ref{sec:rate}, we present stronger evidence on this statement, through modeling the BCG redshift evolution for both observational data and simulation data, testing the model and eventually showing the model constraints being different in the observation and in the simulation.

In addition, we are not considering BCG luminosity and stellar mass uncertainties in this section (they are not included in the linear fitting procedure). We also address this in Section ~\ref{sec:shm} and~\ref{sec:rate}.

\subsection{High Redshift BCGs}
\label{sec:hiz}

At $z>0.9$, we notice that two of the four BCGs in our sample appear to be massive/luminous outliers by $\sim$ 0.5 dex, which matches previous findings about massive BCGs at z > 1.0. In \citet{2009Natur.458..603C},  five $1.2<z<1.5$ BCGs are identified to be $0.5\sim0.7$ dex more massive than the DL07 simulation BCGs, and in \cite{2013ApJ...769..147L}, a massive $z=1.096$ cD type galaxy is discovered in a 5 $\mathrm{arcmin^2}$ {\it Hubble} Deep Field. However, after considering cluster mass uncertainty, and the BCG luminosity and stellar mass uncertainties, we can only detect the over-massive/over-luminous BCG effect with $\sim$ 1 $\sigma$ significance. 

\section{BCG-Cluster Mass Relation}
\label{sec:shm}

To further investigate the growth of BCGs, we turn to modeling a redshift-dependent, stellar-to-halo mass relation. We refer to this relation as the BCG-Cluster mass relation in this paper. Later, in Section~\ref{sec:rate}, we use this relation to model the BCG growth rate from z = 1.0 to z = 0.

\begin{figure*}
\includegraphics[width=1.0\textwidth]{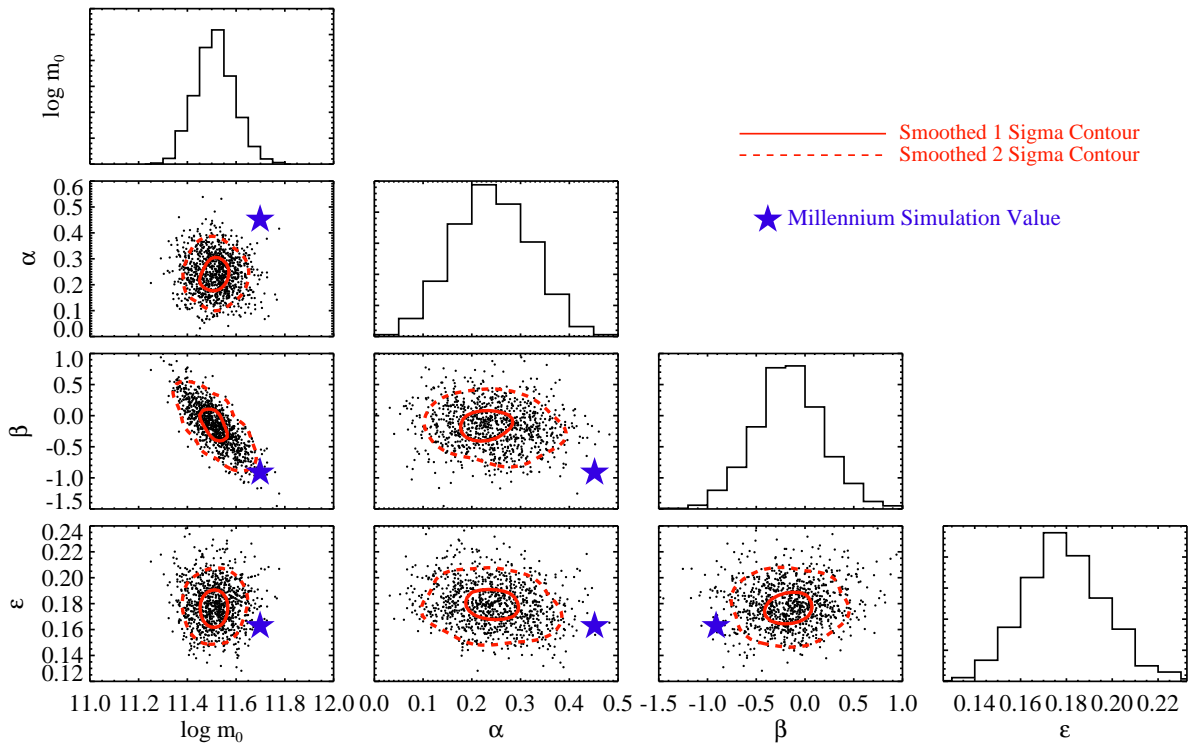}
\caption{Posterior distribution of the parameters, $\mathrm{log} m_0$, $\alpha$, $\beta$ and $\epsilon$, in the BCG-cluster mass relation (Equation~\ref{eq:shm}), based on the BCG stellar mass derived in 32 kpc aperture. The histogram in each column shows the marginalized distribution of the corresponding parameter. Other panels show the correlation between two parameters noted on the x and y axes.}
\label{fig:mass_params}
\end{figure*}

\begin{table*}
\begin{center}
\caption{Constraints on $\mathrm{log}m_0$, $\alpha$, $\beta$ and $\sigma$ of the BCG-cluster mass relation (Equation~\ref{eq:shm}) and $p$ from the outlier pruning procedure \label{tbl:shm}}
\begin{tabular}{ccccccccc}
\tableline\tableline
   & Prior  & 15 kpc & 32 kpc & 50 kpc & 60 kpc & 32 kpc ($\mathrm{log} M_{200}$ > 13.85) &  32 kpc ($\mathrm{log} M_{200}$ <  14.70) & DL07 \\
 \tableline
$\mathrm{log} m_0$ &  [10, 13]  & $11.37\pm0.08$ & $11.52 \pm 0.08$ & $11.60\pm 0.09$ & $11.61\pm0.09$& $ 11.58 \pm 0.08 $ & $  11.49 \pm 0.08$  &$ 11.698 \pm 0.004 $\\
\tableline
$\alpha$ & [-0.5, 0.8]  & $0.20\pm0.08$ & $0.24\pm 0.08$ & $0.30\pm0.08$ & $0.32\pm0.09$ & $ 0.37 \pm 0.10 $ & $0.19\pm0.11$ & $0.452\pm0.004$\\
\tableline
$\beta$ & [-2, 2] & $-0.15\pm0.31$ & $-0.19\pm0.34$ & $-0.24\pm0.39$ & $-0.19\pm0.40$ & $ -0.62 \pm 0.34 $& $-0.06\pm0.37$& $-0.912\pm 0.026$\\
\tableline
$\epsilon$ & [0.001,1] & $0.172\pm0.015$& $0.180\pm0.018$ & $0.192\pm0.019$ & $0.198\pm0.020$ & $ 0.169 \pm 0.019 $ & $0.186\pm0.020$ & $0.1628\pm0.0012$\\
\tableline
$p$ & [0.5, 1.0] & $0.970\pm0.017$& $0.970\pm0.018$ & $0.971\pm0.017$ & $0.970\pm0.018$ & $ 0.965 \pm 0.019 $ & $0.967\pm0.017$ & N.A. \\
\tableline
\end{tabular}
\end{center}
\end{table*}

\subsection{Modeling the BCG-Cluster Mass Relation}
\label{sec:compt_shm}

We model the BCG-Cluster Mass Relation as redshift dependent with the following equation,
\begin{equation}
\mathrm{log} m_{*}=\mathrm{log} m_{0}+{\alpha}\mathrm{log} (\frac{M_{200, ~z}}{\mathrm{M_{piv}}})+\beta \mathrm{log}(1+z).
\label{eq:shm}
\end{equation}

This equation adopts a power law dependence on cluster mass \citep{2014MNRAS.440..762O, 2014arXiv1401.7329K, 2008MNRAS.385L.103B, 2010ApJ...710..903M, 2013MNRAS.428.3121M} as well as a power law dependence on redshift. We choose $\mathrm{M_{piv}}$ to be $1.5\times10^{14}M_{\odot}$, about the median mass of the XCS clusters. We also assume that there exists an intrinsic scatter, $\epsilon$, between the observed BCG stellar mass and this relation, as $\mathrm{log} m_{*, obs} \sim \mathcal{N} (\mathrm{log} m_{*}, \epsilon^2)$. Hence, the relation contains four free parameters: $\mathrm{log}m_0$, $\alpha$, $\beta$ and $\epsilon$. 

We perform a Markov Chain Monte Carlo (MCMC) analysis to sample from the following posterior likelihood:

\begin{equation}
  \mathrm{log}\mathcal{L} =-\frac{1}{2} \mathrm{log} | {\bf C}|  -\frac{1}{2} {\bf Y^{\mathrm{T}} C^{-1} Y } +\mathrm{log}p(\mathrm{\bf Q}).
\label{eq:likelihood}
\end{equation}
In this function, $\mathrm{\bf Y}$ is a 106 dimension vector $(y_1, y_2 ..., y_{106} )$, with the $k$th element being the difference between the modeled and the observed BCG stellar masses, as:
\begin{equation}
y_k = y_\mathrm{model, k} - y_\mathrm{obs, k}.
\end{equation}

The covariance matrix, $\mathrm{\bf C}$, in Equation~\ref{eq:likelihood} is the combination of the covariance matrices for cluster mass measurements, BCG stellar mass measurements, redshift measurements and the  intrinsic scatter. It has the following form:

\begin{equation}
{\bf C} = {\bf Cov(m_{*})+\alpha^2Cov (\mathrm{log}M_{200}) } {\bf + \beta^2 Cov(\mathrm{log}(\textnormal{1}+z)) } + \epsilon ^2 {\bf I}. 
\label{eq:covariance}
\end{equation}

Additionally, we implement an outlier pruning procedure as we ``fit'' (or sampling the posterior distribution in Bayesian statistics) for the BCG-cluster mass relation, as described in \cite{2010arXiv1008.4686H}. To summarize this procedure, we adopt a set of binary integers $\mathrm{{\bf Q}=(q_1, q_2, ..., q_{106})}$ as flags of outliers. $\mathrm{q}_{k}=0$ indicates an outlier and $y_k$ is correspondingly modified as, 
\begin{equation}
y_k=\mathrm{log}m_{*, k}-\mathrm{log}m_\mathrm{outlier}, 
\end{equation}
where $m_\mathrm{outlier}$ is treated as a 5{\it th} free parameter. To penalize data pruning, we assume a Bernoulli prior distribution for $\mathrm{{\bf Q}}$, characterized by another free parameter $p$ as,
\begin{equation}
p(\mathrm{\bf Q})=\prod_{k}p^{q_k}(1-p)^{1-q_k}.
 \end{equation}
 
Eventually, the parameters to be sampled from Equation~\ref{eq:likelihood} are $\mathrm{log} m_0$, $\alpha$, $\beta$, $\epsilon$, $\mathrm{\bf Q}$, $p$, and $\mathrm{log}m_\mathrm{outlier}$. More details about deriving the posterior likelihood (Equation~\ref{eq:likelihood}) as well as choosing the covariance matrix  can be found in Appendix~\ref{sec:additional_likelihood}. We assume uniform truncated priors for all the free parameters except $\bf Q$, and the final result appears to be insensitive to this choice. We perform the fitting procedure for both the observed BCGs from the XCS sample and the simulation BCGs sampled from the DL07 simulation (Section~\ref{sec:sim_select}).

\subsection{Constraints on the BCG-Cluster Mass Relation}
\label{sec:constraint_shm}

%\begin{figure*}
%\includegraphics[width=1.0\textwidth]{mass_model_panel.eps}
%\caption{(a, b, c, d) Differences between the observed and modeled BCG stellar masses, plotted against redshift and cluster mass. (e, f) Re-presentation of the differences between observed and simulated BCG stellar masses (see Figure~\ref{fig:mass_sim}). BCGs in low mass clusters (bottom row) appear to be over-massive outliers at 2 (in panel b) to 3 (in panel d,f) $\sigma$ significance levels in these threee comparisons. We also mention it here that our redshift-dependent BCG-cluster mass relations (a, b, c, d) fit the data better than the DL07 simulation (e, f).}
%\label{fig:masscut_panel}
%\end{figure*}

In Figure~\ref{fig:mass_params}, we plot the posterior distribution of $\mathrm{log} m_0$, $\alpha$, $\beta$, $\epsilon$ in Equation~\ref{eq:shm}. We also list their marginalized means and standard deviations in Table \ref{tbl:shm}.

The constraint we derive on $\alpha$ agrees well with the reported values from the literature \citep{2014MNRAS.440..762O, 2014arXiv1401.7329K, 2008MNRAS.385L.103B}. We also notice that $\alpha$ increases with bigger BCG apertures, indicating stronger correlation with cluster mass in the BCG outskirts \citep[also see][]{2012MNRAS.422.2213S}. This effect seems to be justifiable, considering an inside-out growth scenario for BCGs \citep{2010ApJ...709.1018V, 2013ApJ...766...15P, 2014ApJ...789..134B}. Further analysis with large apertures is limited by the increasing amount of background noise at BCG outskirt, but a larger BCG sample may help quantifying the effect. This effect also illustrates the importance of understanding BCG photometry measurement when deriving BCG-cluster mass relations.

Our estimation of $\mathrm{log} m_0$, the normalization of  Equation~\ref{eq:shm}, appears to be lower than the corresponding value in DL07 by 0.1 -- 0.2. As $\mathrm{log} m_0$ is mainly constrained by low redshift BCGs, this result is completely consistent with BCGs being under-massive at low redshift as discussed in Section~\ref{sec:loz}.

Our estimation of $\beta$, the index of the redshift component in Equation~\ref{eq:shm}, also disagrees with the corresponding value in DL07. The constraint on $\beta$ derived from the whole cluster sample is different from the simulation value at a significance level of  2.3 $\sigma$. The constraint from our data is closer to 0, suggesting less change of BCG stellar mass with redshift. Note that a further, quantitative conclusion should not be drawn. Although $\beta$ is the dominant parameter that describes BCG redshift evolution in Equation~\ref{eq:shm}, it is not the only one. The mass term  in Equation~\ref{eq:shm} also contains information about BCG redshift evolution as cluster $M_{200}$ evolves with time. A quantitative analysis of BCG redshift evolution is presented in Section~\ref{sec:rate}.

Our constraint on $\beta$ is highly co-variant with $\mathrm{log} m_*$ (recall the bi-variate normal distribution), but the co-variance shall not be interpreted as ``degeneracy'': a reasonable $m_*$ sampled from its marginalized posterior distribution does not make $\beta$ consistent with the simulation. We also notice that different conventions for BCG magnitude measurement can bias the constraint on $\beta$. For example, using the Kron magnitude from the popular SExtractor software \citep{1996A&AS..117..393B}, which tends to under-estimate BCG Kron Radius and therefore BCG total magnitude (See discussion in Section~\ref{sec:kron}. This effect happens frequently for our intermediate redshift BCGs), shifts $\beta$ downward by $\sim$ 1 $\sigma$.

We detect hints that the constraints on $\alpha$ and $\beta$ may depend on cluster mass (see Table~\ref{tbl:shm}). For clusters with $\mathrm{log} M_{200}$ above $13.85$, we notice stronger correlation between BCG and cluster masses \citep[larger $\alpha$, compare it to ][]{2014arXiv1412.7823C,2014A&A...561A..79V} and steeper redshift evolution (smaller $\beta$) at $\sim$ 1.0 $\sigma$ significance level. However, 
%inspection (see Figure~\ref{fig:masscut_panel}) shows
BCGs in low mass clusters ($\mathrm{log} M_{200}$ < 13.85) are possibly over-massive compared to our simulation calibrated BCG-cluster mass relation.
%outliers to our relations with $\sim$ 2 $\sigma$ significance. 
Evaluating the masses of low mass clusters and groups through their X-ray observables needs to be handled with care. In this paper, we use lensing calibrated $M-T$ relation of galaxy groups and clusters to derive $M_{200}$ for most low mass clusters (see Figure~\ref{fig:lm} and Section~\ref{sec:mass}). Arguably, the accuracy of X-ray inferred masses of low mass clusters is less well characterized than the higher mass end. Thus, in our growth rate determinations we show the difference after excluding the lowest mass systems ($\mathrm{log} M_{200}$ < 13.85, about 10\% of the sample). For consistency, we also examine the effect of excluding the highest mass systems ($\mathrm{log} M_{200}$ > 14.7, about 10\% of the sample).

\section{BCG Stellar Mass Growth since $z\sim1.0$}
\label{sec:rate}

\begin{figure}
\includegraphics[width=0.5\textwidth]{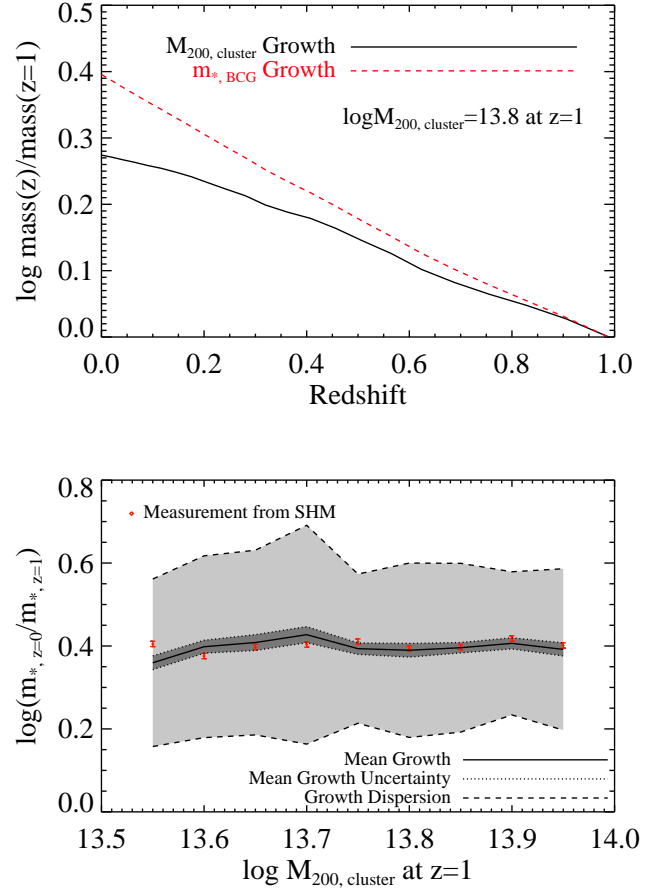}
\caption{ To derive BCG stellar mass growth rate from BCG-cluster mass relations, we will need to derive cluster mass growth history from simulations. In the top panel, we show the halo mass evolution history of $\sim200$ halos with $\mathrm{log} M_{200}=13.8$ at $z=1.0$. We utilize Equation~\ref{eq:growth} (we are using the BCG-cluster mass relation in DL07) to derive the BCG stellar mass growth rate shown by the red dashed line. In the bottom panel, we show the test result for this method (see Section~\ref{sec:rate_comp} for details). Overall, our approach well reproduces the average growth rate within $1\sigma$ for simulation BCGs. }
\label{fig:rate_method}
\end{figure}

In this section, we compute the BCG stellar mass growth rate since redshift $1.0$. We derive the growth rate conveniently using the redshift-dependent BCG-cluster mass relation from the previous section. Doing so, we are assuming that a redshift-dependent BCG-cluster mass relation not only describes the relation between BCG stellar mass and cluster mass at various redshifts, but also describes how BCG stellar mass evolves with time. There is no new measurement made with observational data in this section. The redshift-dependent BCG-cluster mass relation derived in the previous section is the only input from observational data. Our method, however, do need new input from simulation data, which is the mass evolution history of clusters.

In this section, we compute the stellar mass growth rate for the BCGs hosted by clusters of $\mathrm{log} \,M_{200}=13.8$ at $z=1.0$. The choice is made as the XCS sample well represents these clusters and their low redshift descendants (see Figure~\ref{fig:rate_method} for the mass evolution history of clusters with $\mathrm{log} \,M_{200}=13.8$ at $z=1.0$). The method is also applied to clusters of different masses, but we do not notice significant change of the conclusions.

\subsection{Method and Test}
\label{sec:rate_comp}

We need to know how the cluster mass evolves with redshift in our method.
To acquire this information, we select a sample of halos with $z\sim1.0$, $\mathrm{log} \,M_{200}\sim13.8$ from the Millennium simulation, and extract their evolution history by identifying descendants of these halos all the way to $z=0$ (using the {\it descendantid} keyword).  We then compute the mean $M_{200}$  evolution of these halos, shown in Figure~\ref{fig:rate_method}. 

The second step is to use the BCG-cluster mass relation to derive the average stellar mass of the BCGs hosted by these halos at different redshifts.
% Simple but not obvious, 
From Equation~\ref{eq:shm}, the average BCG stellar mass relative to some normalization epoch, $z_0$, can be expressed as:
\begin{equation}
\mathrm{log} \frac{m_{*, z}}{m_{*, z_0}}=\alpha \mathrm{log} \frac{M_{200, z}}{M_{200, z_0}} + \beta \mathrm{log}\frac{1+z}{1+z_0}.
\label{eq:growth}
\end{equation}
We take $\mathrm{log} \frac{m_{*, z}}{m_{*, z_0}}$ from the above equation as describing the average BCG stellar mass growth. The $\frac{M_{200, z}}{M_{200, z_0}}$ component in the equation is the average cluster mass growth extracted from the simulation.

The result of applying Equation~\ref{eq:growth} to the average  $M_{200}$ growth with the simulation data is also shown in Figure~\ref{fig:rate_method}. We estimate the uncertainties on the BCG stellar mass growth rate through sampling the joint constraint on $\alpha$ and $\beta$. We do not consider the uncertainties of cluster mass evolution as it is marginal and is cosmology dependent.

We test our method by applying it to the DL07 simulation BCGs. We first derive the BCG-cluster mass relation in DL07 using the procedure from Section~\ref{sec:compt_shm} for the sample drawn from Section~\ref{sec:sim_select}. We compare the computed BCG growth rate to the values obtained through directly tracking cluster descendants. The latter is acquired through recording the central galaxy stellar mass of the halo descendants since redshift 1.0. We consider the result from this second approach as the {\it true} growth of simulation BCGs. 

In the bottom panel of Figure \ref{fig:rate_method}, we show the BCG growth rate derived with Equation~\ref{eq:shm}, and the {\it true} growth rate encompassed by uncertainty from bootstrapping. Overall, for low mass clusters, our approach reproduces the average BCG growth rate from z = 1.0 to z =0 within $1\sigma$. Bias associated with this method \citep[like progenitor bias, see:][]{2015arXiv150102800S}, if there is any, appears to be negligible.

\subsection{Growth Rate from z = 1.0 to z = 0}
\label{sec:growth_result}

\begin{figure}
\includegraphics[width=0.5\textwidth]{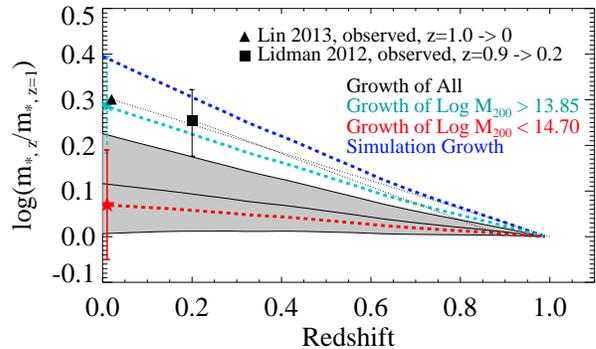}
\caption{This plot shows our BCG stellar mass growth estimation with the full sample and with two mass-limited sub-samples. We also show the measurements in \cite{2012MNRAS.427..550L, 2013ApJ...771...61L} and the BCG growth rate in the DL07 simulation. Our estimation is consistent with previous measurements, but slower than DL07 by $ \sim$ 2.5 $\sigma$. The uncertainty from DL07 is extremely small as the simulation is well sampled (see Figure~\ref{fig:rate_method}).}
\label{fig:rate_res}
\end{figure}

We compute the BCG stellar mass growth rate using Equation~\ref{eq:growth} and compare it to the simulation value obtained with the same method. We discuss the observational result  based on BCG 32 kpc aperture stellar masses in this paper -- the result derived with other apertures look similar. From z = 1.0 to z = 0, we estimate the BCG growth rate to be $0.13\pm0.11$ dex, comparing to $0.40\pm 0.05$ dex in simulation (uncertainty estimated for the BCG sample in Section~\ref{sec:sim_select}), as shown in Figure~\ref{fig:rate_res}. This result is in agreement with our conclusion from the simulation matching analysis (Section~\ref{sec:loz}), and also in agreement with previous studies \citep{2012MNRAS.427..550L, 2013ApJ...771...61L}. Even after considering all the uncertainties, biases and covariances associated with BCG luminosity and stellar mass measurements, we still confirm that the observed BCG growth is slower than the prediction from DL07 at a significance level of $\sim$ 2.5 $\sigma$.

Like our constraint on the BCG-cluster mass relation, our result here shifts by $\sim$ 1 $\sigma$ ($0.29\pm0.11$ dex) when we exclude the lowest mass systems ($\mathrm{log} M_{200}$ < 13.85, about 10\% of the sample). Note that the shift may be caused by inaccuracy of X-ray cluster mass scaling relations at the low mass end (see discussion in Section~\ref{sec:constraint_shm}). For consistency, we also show the result ($0.07\pm0.12$ dex) after excluding the highest mass systems ($\mathrm{log} M_{200}$ > 14.7, about 10\% of the sample). Our result is also susceptible to improper BCG magnitude measurements. Using the Kron magnitude from SExtractor, the result will be biased toward more rapid BCG growth by $\sim$ 1 $\sigma$. We also considered applying our method with stellar-to-halo mass relations from literature, but as many previous studies are based on magnitude conventions with various problems for BCGs (see discussion in Appendix~\ref{sec:phot}), we opt for not using them in this paper.

\subsection{Role of ICL Production}

\begin{figure}
\includegraphics[width=0.5\textwidth]{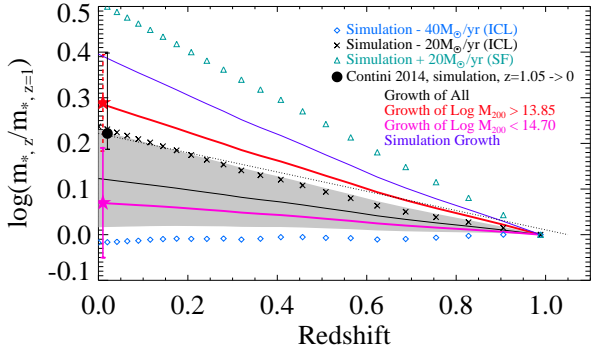}
\caption{We show BCG growth rates from toy models adding more rapid ICL production or more dramatic BCG star formation to the DL07 simulation. Including extra constant ICL production at 20 - 40 $M_{\odot}/\mathrm{yr}$ well reproduces our observed growth rate. We also show the predicted BCG growth rate from \cite{2014MNRAS.437.3787C}, which has updated the DL07 simulation with more realistic ICL production. The BCG growth rate in \cite{2014MNRAS.437.3787C} agrees well with our measurement.}
\label{fig:bcg_icl}
\end{figure}

In this paper, we have shown from two different perspectives that the BCG stellar mass growth rate in clusters with $\mathrm{log} \,M_{200}=13.8$ at $z=1.0$ is slower than the prediction naively expected in a hierarchical formation scenario \citep{2007MNRAS.375....2D}. This effect is not that surprising with a second thought on the processes that contribute to (or counter-act) BCG formation.

A hierarchical structure formation scenario predicts that galaxy mergers add stars to BCGs. The BCG stellar build-up can be further augmented by {\it in situ} star formation, but a reduction in stellar mass is possible from mergers that eject stars into the intracluster space. The competition between these mechanisms remains a subject to large modeling uncertainties in simulations. If we assume that BCGs experience the rapid build-up events (mostly merging events) as prescribed in the DL07 simulation, there must be a mechanism that offsets BCG growth to mimic slower evolution we observe in this paper. 

In Figure~\ref{fig:bcg_icl}, we experiment with incorporating extra stellar mass gain or loss into the DL07 simulation. Stellar mass gain tends to steepen BCG growth over time, while stellar mass loss tends to slow down BCG build-up and flatten the BCG growth curve. In order to explain the observed BCG growth rate in our data, the BCGs in DL07 would need to go through extra stellar mass loss at 20 - 40 $M_{\odot}\mathrm{/yr}$, ending up with 2.0-3.5 $\times10^{11}\,M_{\odot}$ at z = 0, which agrees with our data.

Such stellar content stripping from mergers would produce intra-cluster light (ICL). Our result indicates that ICL accumulates at 20-40 $M_{\odot}/\mathrm{yr}$ after $z=1.0$, totaling $(1.5-3)\times10^{11}M_{\odot}$ ICL in the present epoch. This amount corresponds to about 30\% - 60 \% the total of BCG and ICL stellar masses, consistent with the observed ICL fraction in low and medium redshift clusters \citep[$z< 0.5$: ][]{2005MNRAS.358..949Z, 2006AJ....131..168K, 2007AJ....134..466K, 2007ApJ...666..147G, 2011MNRAS.414..602T, 2014ApJ...794..137M, 2014ApJ...781...24G, 2014A&A...565A.126P}. 

In fact, ICL production has already been suggested as an explanation to the seemingly mild evolution of massive galaxies \citep{2006ApJ...652L..89M, 2007ApJ...668..826C, 2012MNRAS.425.2058B, 2013ApJ...770...57B, 2014MNRAS.440..762O}. Although not completely settled, recent studies indicate that ICL possibly forms late, mostly after z=1.0 \citep{2014MNRAS.437.3787C, 2007ApJ...668..826C}. Specifically, the \cite{2014MNRAS.437.3787C} study updates the DL07 simulation with more realistic ICL production processes, and predicts slower BCG growth rate (Figure~\ref{fig:bcg_icl}), in excellent agreement with our measurement. Hence, the slow BCG stellar mass growth since z = 1.0 observed throughout this paper is completely justifiable if ICL forms late after $z=1.0$.

Admittedly, the DL07 simulation also includes stellar stripping that would produce ICL. Unfortunately the amount of ICL from this simulation is not retrievable, and we are not able to analyze if it meets our expectation. The \cite{2011MNRAS.413..101G} SAM simulation has explicitly included ICL production and predicts very similar BCG growth with DL07, but much of the ICL is already in place before $z=1.0$, which is not favored in our interpretation.

\section{Summary and Discussion}
\label{sec:summary}

Using new photometric data from DES and a new X-ray selected cluster and group sample from the XCS, we investigate the redshift evolution of BCG stellar mass since $z=1.2$. We derive constraints on the BCG-cluster mass relation,  and compute the BCG stellar mass growth rate for our sample. From two different perspectives, we demonstrate that the BCG stellar mass growth since z = 1.0 is slower than the expectation from a semi-analytical simulation implementing a simple hierarchical BCG formation scenario. The discrepancy is detected with a significance level as high as 2.5 $\sigma$. We find this slow growth rate after z = 1.0 to be compatible with the late formation of ICL .

We have carefully considered various uncertainties related to studying BCG growth in this work, including the uncertainties of BCG stellar mass measurements, cluster/BCG redshift measurements and cluster mass measurements. We explicitly consider these uncertainties through likelihood analysis, and expect this analysis to help clarify ongoing discussions about how statistical and systematic uncertainties affecting BCG growth measurements.

We also adopt a simple but novel method to compute BCG stellar mass growth rate. Despite considerable attention paid to this topic in the literature, BCG stellar mass growth has been studied with various techniques inconsistent with each other. Ideally, one would like to evaluate BCG stellar mass growth by comparing the BCG masses within the same cluster at high and low redshifts, as we did for method testing in Section~\ref{sec:rate_comp}. This is not possible with observations. However, \cite{2013ApJ...771...61L} have adopted the idea through constructing a cluster sample that resembles the average halo evolution history. In observational studies, the more common approach is to compare the BCG masses of a high redshift cluster sample and a low redshift cluster sample, while adjusting the cluster mass binning at different redshifts to account for cluster mass evolution \citep{2009Natur.458..603C, 2012MNRAS.427..550L, 2014ApJ...789..134B}. The results from these observational studies are widely compared to \cite{2007MNRAS.375....2D}, which computes the BCG stellar mass growth rate through a ``fixed space density'' method, i.e, selecting  the 125 most massive clusters at $z\sim1.0$ and $z\sim0$ respectively to  compare their BCG masses. Compared to these previous studies, our method allows consistent comparison to simulation for clusters of specific masses and redshifts. Our test in Section~\ref{sec:rate_comp} shows that the approach suffers from only negligible bias for the required precision.

Finally, the analyses presented in this paper are based on DES SV data, a data set corresponding to only 5\% of the nominal DES footprint. With spectroscopic and X-ray follow-up, Miller et al. (in prep) show that the final DES/XCS sample should be about 10 times larger than this data set. Comparing the constraints on the BCG-cluster mass relation derived with 1000 simulation clusters rather than 100 of them, we conclude that we expect $\sim$ 3 times improvement in the measurement uncertainty of BCG growth. 
At this level of statistical power, it will be critically important to thoroughly understand the uncertainties associated with various observables. This paper presents the first steps toward such an analysis.

\acknowledgements

The authors are pleased to acknowledge support from the University of Michigan Rackham Predoctoral Fellowship and Department of Energy research grant DE-FG02-95ER40899. We are also indebted to Eric Bell, Dragan Huterer, Gabriella De Lucia, Heidi Wu, Mariangela Bernardi for helpful discussions. We thank the anonymous referee for a very careful reading of the paper and the many helpful suggestions. 

We use DES science verification data for this work. We are grateful for the extraordinary contributions of our CTIO colleagues and the DES Camera, Commissioning and Science Verification teams in achieving the excellent instrument and telescope conditions that have made this work possible. The success of this project also relies critically on the expertise and dedication of the DES Data Management organization. 

Funding for the DES Projects has been provided by the U.S. Department of Energy, the U.S. National Science Foundation, the Ministry of Science and Education of Spain, 
the Science and Technology Facilities Council of the United Kingdom, the Higher Education Funding Council for England, the National Center for Supercomputing 
Applications at the University of Illinois at Urbana-Champaign, the Kavli Institute of Cosmological Physics at the University of Chicago, 
the Center for Cosmology and Astro-Particle Physics at the Ohio State University,
the Mitchell Institute for Fundamental Physics and Astronomy at Texas A\&M University, Financiadora de Estudos e Projetos, 
Funda{\c c}{\~a}o Carlos Chagas Filho de Amparo {\`a} Pesquisa do Estado do Rio de Janeiro, Conselho Nacional de Desenvolvimento Cient{\'i}fico e Tecnol{\'o}gico and 
the Minist{\'e}rio da Ci{\^e}ncia e Tecnologia, the Deutsche Forschungsgemeinschaft and the Collaborating Institutions in the Dark Energy Survey. 

The DES data management system is supported by the National Science Foundation under Grant Number AST-1138766.
The DES participants from Spanish institutions are partially supported by MINECO under grants AYA2012-39559, ESP2013-48274, FPA2013-47986, and Centro de Excelencia Severo Ochoa SEV-2012-0234, 
some of which include ERDF funds from the European Union.

The Collaborating Institutions are Argonne National Laboratory, the University of California at Santa Cruz, the University of Cambridge, Centro de Investigaciones Energeticas, 
Medioambientales y Tecnologicas-Madrid, the University of Chicago, University College London, the DES-Brazil Consortium, the Eidgen{\"o}ssische Technische Hochschule (ETH) Z{\"u}rich, 
Fermi National Accelerator Laboratory, the University of Edinburgh, the University of Illinois at Urbana-Champaign, the Institut de Ciencies de l'Espai (IEEC/CSIC), 
the Institut de Fisica d'Altes Energies, Lawrence Berkeley National Laboratory, the Ludwig-Maximilians Universit{\"a}t and the associated Excellence Cluster Universe, 
the University of Michigan, the National Optical Astronomy Observatory, the University of Nottingham, The Ohio State University, the University of Pennsylvania, the University of Portsmouth, 
SLAC National Accelerator Laboratory, Stanford University, the University of Sussex, and Texas A\&M University.

This paper has gone through internal review by the DES collaboration. It is marked with the DES publication number DES-2015-0040, and the Fermilab preprint number FERMILAB-PUB-15-126-AE.

The Millennium Simulation databases used in this paper and the web application providing online access to them were constructed as part of the activities of the German Astrophysical Virtual Observatory (GAVO). 

\bibliography{references}

\begin{appendix}

\section{Galaxy Cluster Mass}
\label{sec:mass}

\subsection{Cluster Mass From X-ray Temperature}
\label{sec:tmass}

\begin{figure*}
\includegraphics[width=1.0\textwidth]{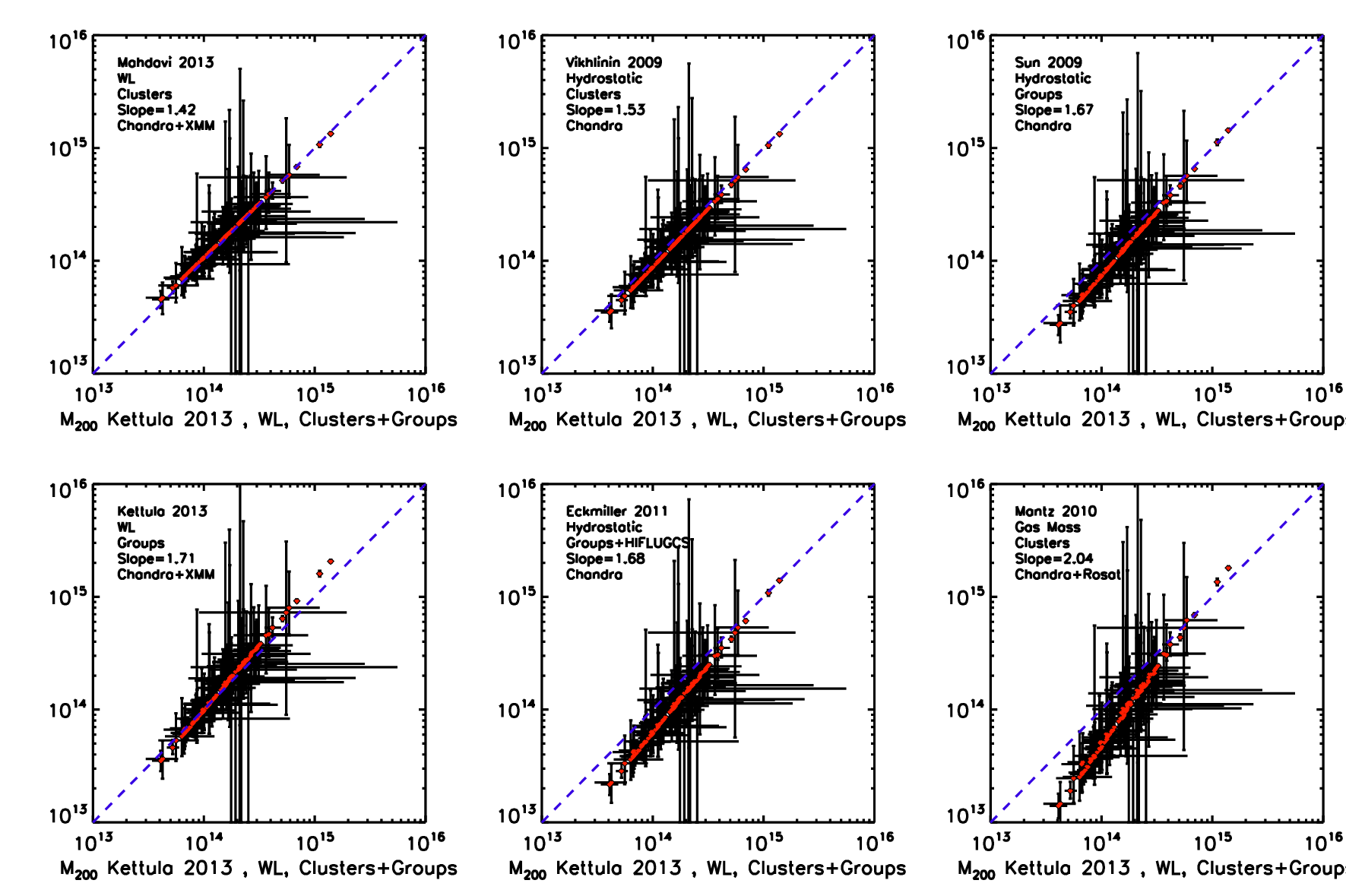}
\caption{Comparison of the cluster masses derived from a few $M-T$ scaling relations. We plot the mass derived from the \cite{2013ApJ...778...74K} relation on the $x-$axis of all the panels. The \cite{2013ApJ...778...74K} $M-T$ scaling relation agrees well with other relations at the cluster scale (See Section~\ref{sec:tmass}). For simplification, we only include X-ray temperature measurement uncertainty in this figure.}
\label{fig:xm}
\end{figure*}

\begin{figure*}
\includegraphics[width=1.0\textwidth]{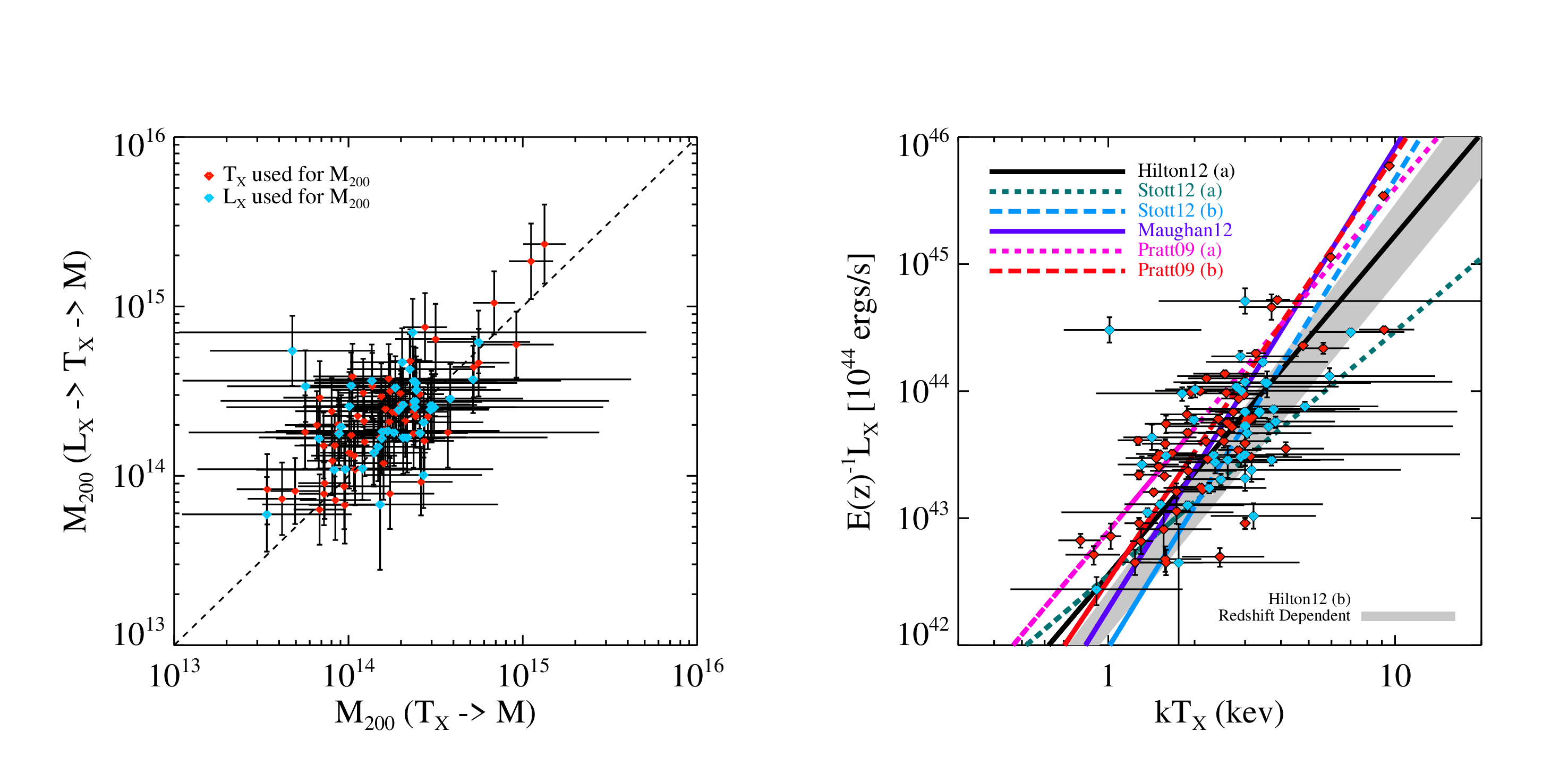}
\caption{Left: We derive cluster masses from both X-ray luminosities and X-ray temperatures, and decide which one to use through comparing their uncertainties. Right: X-ray temperature and X-ray luminosity of the XCS clusters plotted against the scaling relations in \cite{2012MNRAS.424.2086H, 2012MNRAS.422.2213S, 2012MNRAS.421.1583M, 2009A&A...498..361P}.The grey band shows the redshift-dependent $L-T$ relation in \cite{2012MNRAS.424.2086H} between z = 1 and z =0.} 
\label{fig:lm}
\vspace{2em}
\end{figure*}

We use a lensing calibrated $M-T$ scaling relation from \cite{2013ApJ...778...74K} to derive cluster mass from X-ray temperature ($T_{X}$, core not excised). In \cite{2013ApJ...778...74K}, weak lensing mass measurements are obtained for 10 galaxy groups in the mass range $0.3-6.0\times 10^{14}h^{-1}_{70}M_\odot$. Together with 55 galaxy clusters (many above $2\times 10^{14}h^{-1}_{70}M_\odot$) from \cite{2011ApJ...726...48H, 2012MNRAS.427.1298H, 2013ApJ...767..116M}, \cite{2013ApJ...778...74K} derive weak lensing calibrated $M-T$ relation across the group to cluster range \footnote{{\it Chandra} temperatures in \cite{2013ApJ...767..116M} are adjusted to match {\it XMM} calibration.}.

We check the \cite{2013ApJ...778...74K} $M-T$ scaling relation against a few other studies based on the gas content (Figure~\ref{fig:xm}). The Hydrostatic Equilibrium \citep[HSE,][]{2009ApJ...693.1142S, 2011A&A...535A.105E, 2009ApJ...692.1033V} and gas mass fraction \citep{2010MNRAS.406.1773M} calibrated relations agree with the \cite{2013ApJ...778...74K} $M-T$ relation at the cluster scale, but  have troubles matching to it at the group scale \citep{2009ApJ...693.1142S, 2011A&A...535A.105E}. As known from simulations \citep[see \cite{2013ApJ...778...74K} for discussion]{2007ApJ...655...98N, 2012NJPh...14e5018R}, the disagreement is not surprising as HSE masses are biased low at the group scale and the gas mass fraction relation is only derived with the most massive clusters.

We use the  \cite{2013ApJ...778...74K} scaling relation to derive cluster $M_{500}$ from X-ray temperature, and then the \cite{2003ApJ...584..702H} relation to derive $M_{200}$ from $M_{500}$. We assume the cluster concentration parameter to be 5 in this procedure. Using a different concentration parameter in the [3, 5] range only changes $M_{200}$ at percent level. We also assume the intrinsic scatter of $M_{500}$ to be 0.1 dex as typically found in simulation studies \citep{2006ApJ...650..128K, 2009ApJ...691.1307H}.

\subsection{Cluster mass from X-ray Luminosity}

We resort to X-ray luminosity ($L_{X}$, core not excised) to estimate masses for clusters/groups that do not have high quality temperature measurements. We first derive X-ray temperature using a $L-T$ scaling relation, and then derive $M_{200}$ using the procedure above. 

We use a self-similar, redshift-dependent $L-T$ scaling relation from \cite{2012MNRAS.424.2086H}, but also experimented with a few other self-similar $L-T$ relations (see Figure \ref{fig:lm}) from \cite{2012MNRAS.422.2213S, 2012MNRAS.421.1583M, 2009A&A...498..361P}. The \cite{2012MNRAS.424.2086H} relation, which is also based on a XCS sample \citep{2012MNRAS.423.1024M}, provides the best fit to our data. We assume 0.1 dex intrinsic scatter for the derived temperatures, as it is constrained in \cite{2012MNRAS.424.2086H}.

\subsection{Mass Uncertainties and the Choice between Temperature and Luminosity Based Masses}
\label{sec:lmtm}

We decide between  $L_X$ and $T_X$ based masses through comparing their uncertainties. To estimate the mass uncertainties associated with each method, we produce 200 "pseudo-measurements" for each cluster, sampling through temperature/luminosity measurement uncertainty,  the scaling relation uncertainty, and the intrinsic scatter of the relations. We derive mass uncertainties for  $L_X$ or $T_X$ based masses assuming log-normal distribution for the 200 "pseudo-measurements". If the uncertainty of $T_{X}$ mass is larger than the uncertainty of $L_{X}$ mass by 0.05 dex (we prefer $T_X$ mass since $L_X$ mass is more susceptible to biases), we use $L_X$ mass in lieu of $T_X$ mass. In the end, about half of the cluster masses are derived with $L_X$, and most of the clusters masses have < 0.25 dex uncertainty. 

The 200 "pseudo-measurements" are also used to derive the $M_{200}$ covariance between the cluster sample. Because we are including scaling relation uncertainty, the $M_{200}$ covariance matrix is not diagnonal.

\section{BCG Photometry}
\label{sec:phot}

\begin{figure}
\includegraphics[width=1.0\textwidth]{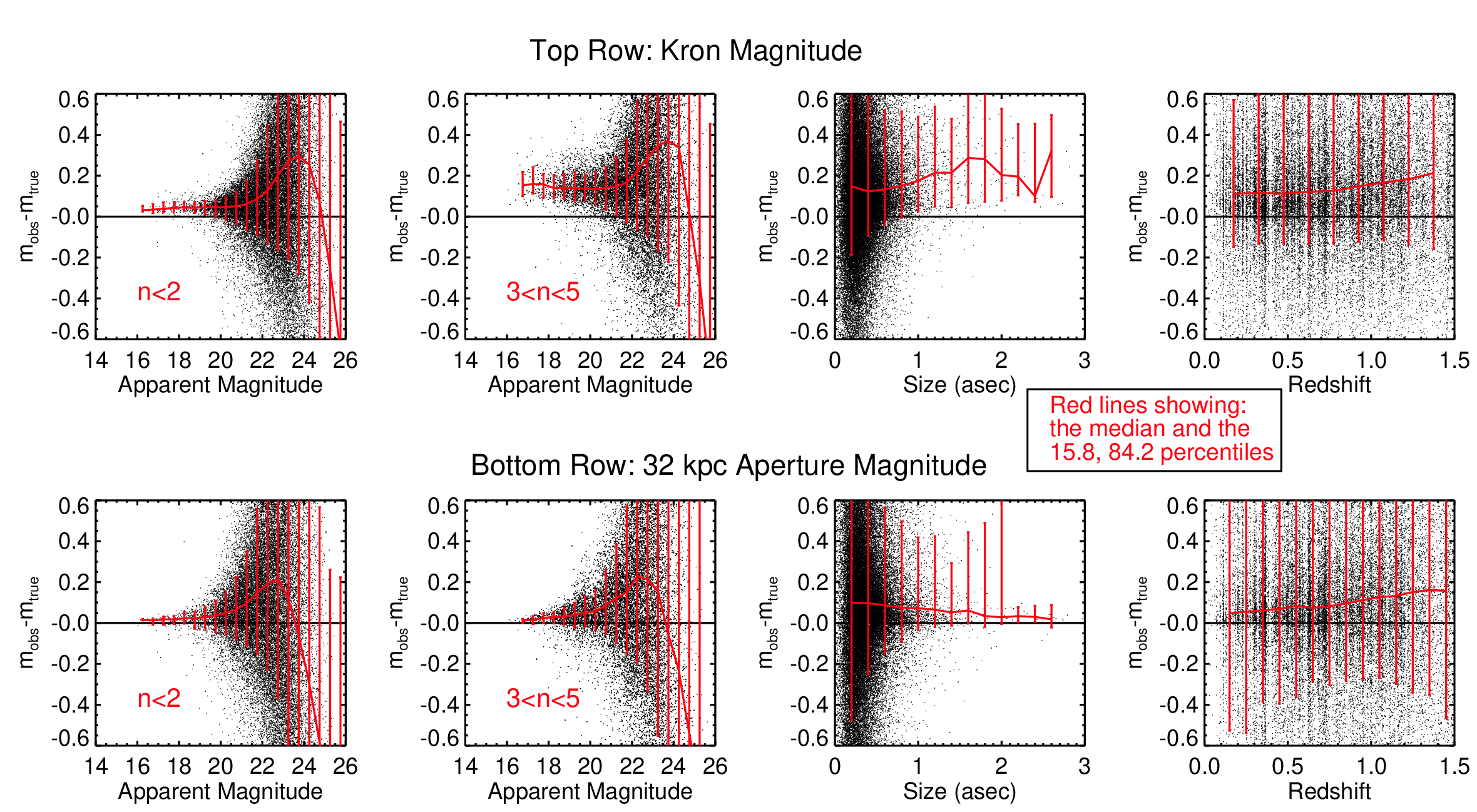}
\caption{We investigate measurement bias associated with Kron magnitude and aperture magnitude using the UFIG sky simulation (see Section~\ref{sec:aperture} for details). Note that this test is done for a general galaxy population rather than BCGs. BCGs below redshift 1.0 generally have Sersic index $>2$ and apparent magnitude below 22. In the top row, $m_{true}$ is the galaxy's input total magnitude, but in the bottom row, $m_{true}$ is the galaxy's input 32 kpc aperture magnitude. To summarize this figure, Kron magnitude tend to under-estimate the brightness of bulge-like galaxies and extended galaxies, while aperture magnitude remain well-behaved for galaxies of all profiles and sizes. The measurements from both systems do become biased for faint galaxies with apparent magnitude above 23, but the bias is un-important for this work. For efficiency, we use SExtractor output in this paper, which compare well with our own measurements (see Appendix~\ref{sec:aperture} for a description on the procedure ) for a general galaxy population.}
\label{fig:test_aper_auto}
\end{figure}

\begin{figure}
\includegraphics[width=1.0\textwidth]{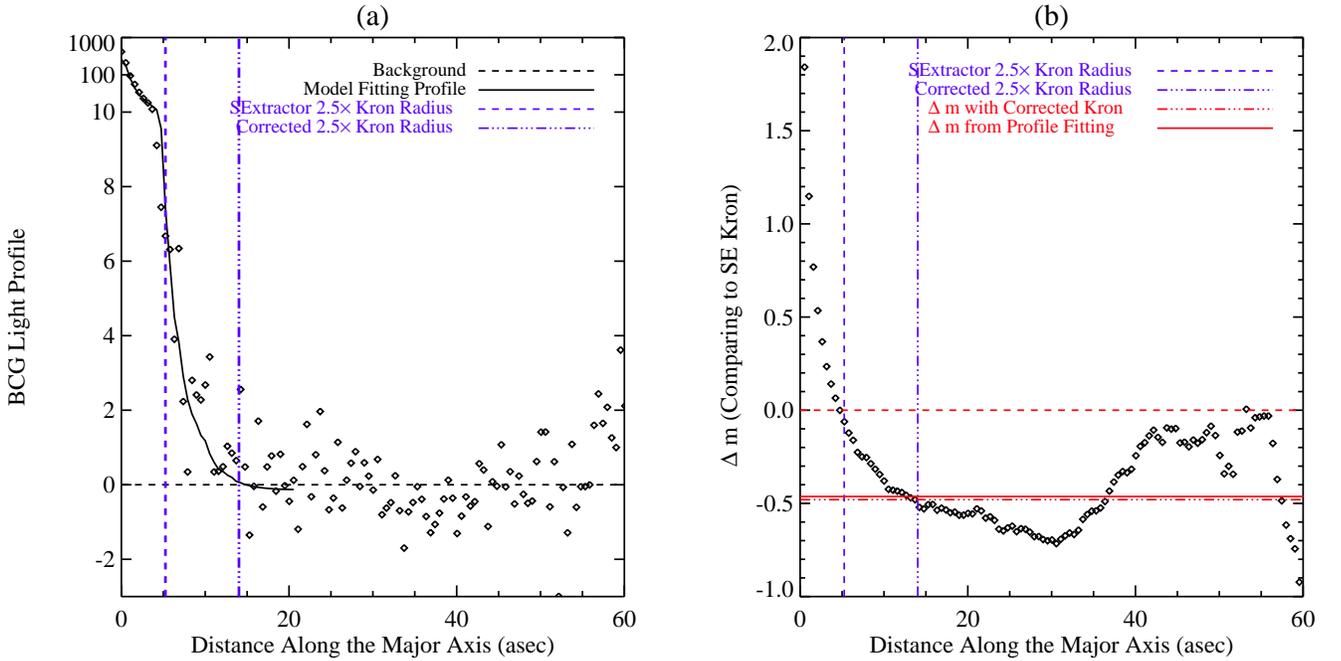}
\caption{The popular SExtractor software tends to under-estimate BCG Kron radius, resulting in significant brightness under-estimation. In panel (b), we show the difference in Kron magnitude measurement, $\Delta m$, when the measurement is made to a different Kron radius ($x$-axis).}
\label{fig:auto_aper_scatter}
\end{figure}

\begin{figure}
\includegraphics[width=1.0\textwidth]{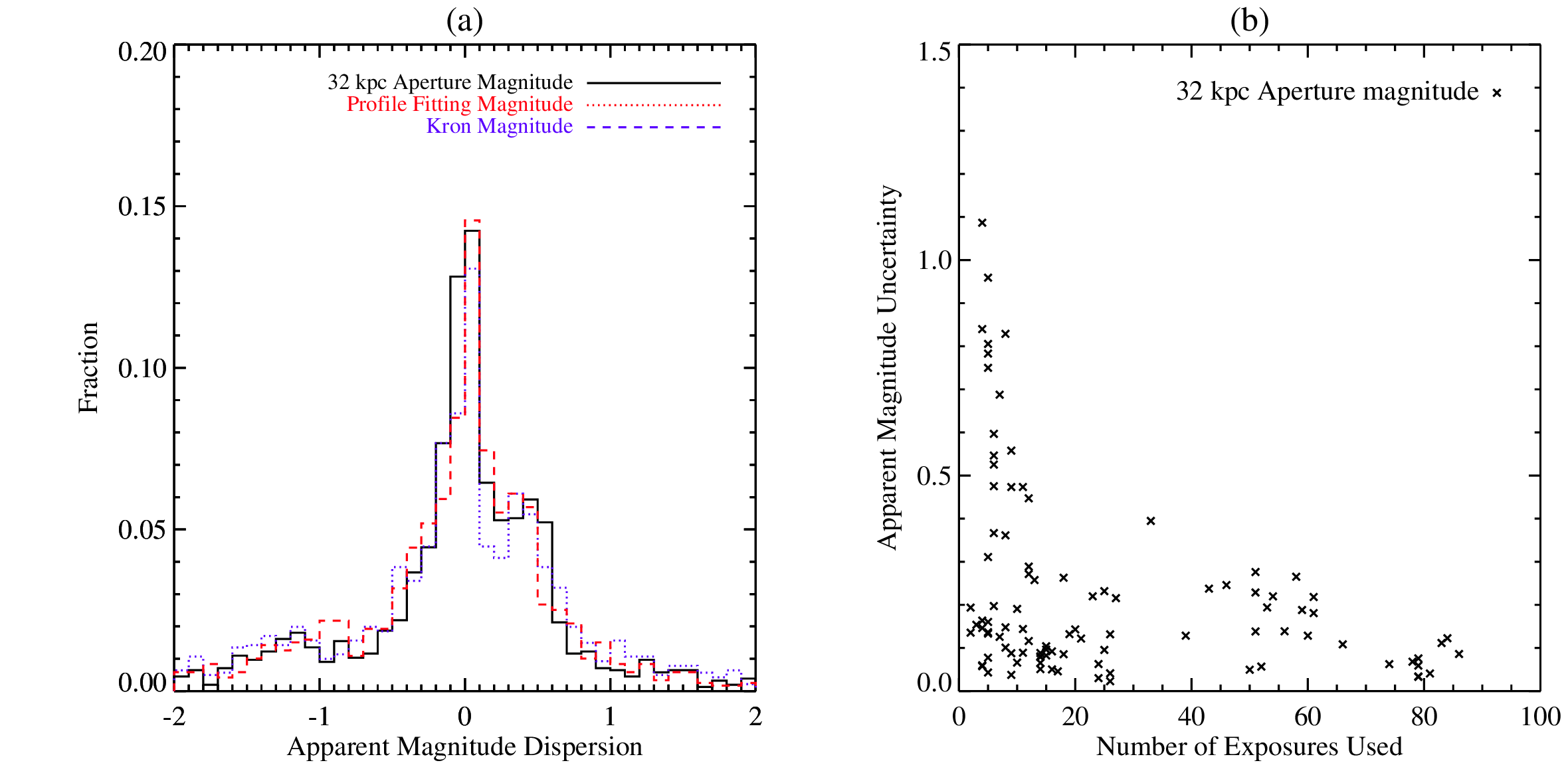}
\caption{(a): Measurement dispersion of aperture magnitude, profile fitting magnitude and Kron magnitude.  (b): We sample BCG apparent magnitude from multiple exposures to evaluate the measurement uncertainty, and our measurement accuracy is limited by the number of exposures we have. }
\label{fig:auto_aper_scatter2}
\end{figure}

BCG photometry measurement is based on products from the official DES Data Management (DESDM) processing pipeline \citep{2012SPIE.8451E..0DM}. In this pipeline, single exposure images are processed, calibrated and later background-subtracted and co-added to produce coadd images. DESDM also runs an advanced version of SExtractor  \citep{2011ASPC..442..435B, 1996A&AS..117..393B} on processed single exposure images and coadd images to produce star/galaxy catalogs, which we do not use because of existing problems for BCGs. In this paper, we derive BCG photometry from processed single exposure images. The following Sections \ref{sec:petrosian} to \ref{sec:aperture} describe our explorations on measuring BCG flux with different magnitude conventions. We discuss potential problems associated with Petrosian magnitude, Kron magnitude, profile fitting magnitude, and aperture magnitude here, but most of the problems are already well analyzed in literature \citep[especially, see][]{2005PASA...22..118G,2007ApJS..172..615H,2007MNRAS.379..867V, 2014MNRAS.443..874B}. Nevertheless, we provide a summary in this Section.

Our final choice is to use BCG photometry measured with 15 kpc, 32 kpc, 50 kpc, and 60 kpc apertures for this paper.

\subsection{Petrosian Magnitude}
\label{sec:petrosian}

Petrosian magnitude measures the flux enclosed within a scaled aperture known as the "Petrosian radius", which is calculated considering background noise level and object light profile \citep{1976ApJ...209L...1P, 2001AJ....121.2358B, 2001AJ....122.1104Y}. It is extraordinarily robust under exposure to exposure variations, but not appropriate for extended galaxies. Although Petrosian magnitude accounts for most of the flux of a disk-like (Sersic index = 1) galaxy, it will only recover 80\% of the flux for  a bulge-like galaxy with a de Vaucouleurs (Sersic index =4) profile \citep{2001AJ....121.2358B}.  

Indeed, a series of studies have found that using Petrosian magnitude \citep[see][for relevant discussion]{2013MNRAS.436..697B, 2013ApJ...773...37H}, the brightness of Luminous Red Galaxies (LRGs) is under-estimated by about 0.3 mag.  Moreover, the missing flux problem is sensitive to the profiles of extended galaxies, and worsens quickly with higher Sersic index. \cite{2005PASA...22..118G} estimate that Petrosian magnitude at its most popular configuration (one that is adopted by SDSS) under-estimates the luminosity of Sersic index = 10 galaxies by 44.7\% (0.643 mag)! For this reason, we are not exploiting Petrosian magnitude in this paper.

\subsection{Kron Magnitude}
\label{sec:kron}

Kron magnitude is another scaled aperture magnitude, measuring the flux enclosed within a few "Kron radius" (usually 2.5 Kron radius), and the Kron radius is decided from the light profile \citep{1980ApJS...43..305K}. Kron magnitude is not as robust as Petrosian magnitude under exposure to exposure variations, but does appear to be more proper for extended galaxies. 

Like Petrosian magnitude, Kron magnitude recovers most of the flux of a disk-like galaxy, but misses 10\% of the flux for a bulge-like galaxy \citep[Sersic index $\sim$ 4,][]{2002A&A...382..495A, 2005PASA...22..118G}. Unlike Petrosian magnitude, the flux missing ratio is in-sensitive to the galaxy Sersic index. \cite{2005PASA...22..118G} estimate that the missing flux varies only at percent level when Sersic index changes from 2 to 10. Indeed, tests with simulated skies \citep[see][ or our test in Figure~\ref{fig:test_aper_auto}]{2002A&A...382..495A} show that Kron magnitude only underestimates the brightness of bulge-like galaxies by about 0.2 mag. It also appears to be indifferent to the presence of ICL: when we apply the measurement to simulated BCGs enclosed by ICL \citep[we use the model  in ][]{2014ApJ...781...24G}, the measurement changes only < $\sim$ 0.1 mag.

As proper as the design of Kron magnitude seems to be, the real problem comes from observationally deriving the Kron radius. As pointed out in \cite{2005PASA...22..118G}, correctly estimating Kron radius requires integration over the light profile to a very large radius, usually many times the half light radius for extended galaxies. If the integration is improperly truncated, the measured Kron radius will be much smaller, and Kron magnitude turns out to be catastrophically wrong -- it may under-estimate the flux of an extended galaxy by as much as 50\% \citep{2002ApJ...571..107B}!  

We find this to be a frequent problem for BCG measurements from the widely-used SExtractor software (i.e., mag\_auto), as demonstrated in Figure~\ref{fig:auto_aper_scatter} (a) and (b). The Kron radius from SExtractor is two times smaller than it should be for one of the BCGs, and the BCG light intensity at  2.5 SExtractor Kron radius is still high. As a result, SExtractor underestimates the Kron flux of this BCG by $\sim$ 0.5 mag. This problem seems purely algorithmic though. Using the galaxy intensity profile to re-calculate Kron radius until it converges, we are able to correct this measurement error. Comparing the corrected measurements to the magnitude measurements from profile fitting (see Section~\ref{sec:profile}), we recover the 0.2 mag accuracy of Kron magnitude as discussed above.

For this paper, we have re-done our analysis using Kron magnitude. We re-compute the Kron radius instead of using SExtractor output, but the result remains qualitatively similar.

\subsection{Profile Fitting Magnitude}
\label{sec:profile}

We have also experimented with BCG profile fitting magnitude from the GALFIT software \citep{2002AJ....124..266P, 2010AJ....139.2097P}. We fit the BCGs w2015PASP..127.1183Zith a model consisting  two Sersic profiles, one with Sersic index = 1 (i.e., a disk profile) and one with flexible Sersic index as suggested by \cite{2014MNRAS.443..874B, 2015MNRAS.446.3943M}. We convolve these models to point spread functions (PSF) derived with the PSFex software \citep{2011ASPC..442..435B}, and carefully mask all neighboring objects including blended objects identified with the GAIN deblender \citep{2015PASP..127.1183Z}. Overall, the design of this procedure is similar to the Galapagos fitting software \citep{2012MNRAS.422..449B}. 

For this paper, we only use the profile fitting magnitude for testing purpose (see Section~\ref{sec:kron} and Section~\ref{sec:aperture}). We hesitate about using it for scientific purpose as we realize that the measurement needs to be extensively tested with sky simulations as in \cite{2007ApJS..172..615H, 2014MNRAS.443..874B, 2015MNRAS.446.3943M}. Upon evaluating the profile fitting magnitude uncertainties (see Figure~\ref{fig:auto_aper_scatter}), we do not find it to improve BCG measurement accuracy and therefore do not consider the testing efforts to be worthwhile for this paper. We nevertheless have re-done our analysis using this magnitude, but the result remains qualitatively similar.

\subsection{Aperture Magnitude Used in this Paper}
\label{sec:aperture}

In this paper, we measure BCG magnitude with circular apertures of 15 kpc, 32 kpc, 50 kpc, and 60 kpc radii. The main results in this paper are derived with the 32 kpc measurements, considering the BCG half light radius measurements in \citep{2011MNRAS.414..445S}. The 32 kpc aperture choice is also comparable to the popular Kron magnitude aperture (2.5 Kron radius) measurements from the SExtractor software. We carefully mask BCG neighbors \citep[including blended objects identified with the GAIN deblender,][]{2015PASP..127.1183Z} and interpolate for the BCG intensity in the masked area. To realistically evaluate BCG magnitude uncertainty, we perform the procedure on processed single exposure images, use the median as the measurement, and evaluate the uncertainty through bootstrapping. We find our typical measurement uncertainty to be $\sim$ 0.4 mag, significantly larger than the SExtractor estimation from co-added images (but not larger when we bootstrap the SExtractor measurements from single exposure images). Since we perform the measurements independently on different exposures, our uncertainty is more comprehensive than the SExtractor uncertainty from co-added images (also see the magnitude measurement scatter test in Figure~\ref{fig:test_aper_auto}). Our measurement becomes uncertain when we have few exposures to work with (see Figure~\ref{fig:auto_aper_scatter}), which will be improved as DES assembles more exposures in the coming years.

To evaluate the sky background level around BCGs, we use background {\it check maps} generated with the SExtractor software from DESDM, configured with the "Global evaluation" process. We sample the values in a ring with inner and outer radius of $\sim$ 13 arcsec and 18 arcsec from the BCG. We have investigated how sky background estimation affects our measurement, as it was considered a difficult task for BCGs. We find it to have only marginal influence.  Indeed, even by using a "Global" setting, SExtractor still over\-estimates the background around some extremely bright sources, known as the {\it dark halo} problem within DES (after background subtraction, the light intensity of a bright object falls slightly below 0 at the outskirt). However, changing the background sampling location only marginally shift our final measurements. In fact, other details of the measurement procedure, like in-complete masking of neighboring sources may cause bigger problems.

We test for measurement bias associated with aperture magnitude and Kron magnitude with simulated DES images, using the UFIG simulation \citep{2013A&C.....1...23B, 2014arXiv1411.0032C}. This sky simulation is based on an N-body dark matter simulation populated with galaxies using the Adding Density Determined GAlaxies to Lightcone Simulations (ADDGALS) algorithm \citep[for a review]{weschlerprep, bushaprep, 2014MNRAS.443.1713D}. We find that both aperture magnitude and Kron magnitude tend to under-estimate the brightness of fainter sources, but the effect is negligible for even the furthest BCGs ($z$ band apparent magnitude is about 22). In addition, the bias would only have suppressed the significance of our result, as further objects are evaluated to be less massive/luminous. We also perform the test with sky simulations based on adding simulated galaxies into real DES co-add images, known as the Balrog simulation \citep{2015arXiv150708336S}, and came to the same conclusion.

\section{BCG Stellar Mass Uncertainty}
\label{sec:bcg_lm}

\begin{table}
\begin{center}
\caption{Parameters of the stellar population models \label{tbl: sps}}
\begin{tabular}{lcc}
\tableline\tableline
Formation Redshift &  20, 10, 8, 5, 4, 3,  2.5, 2, 1.5, 1.0, 0.8, 0.5\\
\tableline
Metallicity & 0.03, 0.025, 0.02, 0.015, 0.01, 0.008, 0.005, 0.003, 0.002\\
\tableline
E-folding time ($\mathrm{Gyrs}$) & 30, 15, 10, 8, 5, 3, 2, 1, 0.8, 0.5, 0.3, 0.1\\
\tableline
Observed Redshift & 1.50, 1.49, 1.48, ..., 0.03, 0.02, 0.01\\
\tableline
\end{tabular}
\end{center}
\end{table}

We derive BCG luminosity and stellar mass with the stellar population modeling technique, and use a SED fitting procedure to find a stellar population model for each BCG. This procedure begins with using the EZGal package \citep{2012PASP..124..606M}, the \cite{2003PASP..115..763C} Initial Mass Function (IMF) and the \cite{2009ApJ...699..486C,  2010ApJ...712..833C} simple stellar population (SSP) models  to produce stellar population templates with various star formation histories and metallicities. We make templates with exponentially-decaying star formation histories (the $\tau$ model) characterized by various e-folding time, metallicity, formation redshift and observed redshift. In Table \ref{tbl: sps}, we list  the parameter values used for these templates.

We then use a Chi-Square minimizing technique \cite[see:][]{2013MNRAS.435...87M} to decide the stellar population template for each BCG. The fitting procedure is done with BCG photometry in DES $g$, $r$, $i$, $z$ bands and we fit the BCGs only to templates of their observed redshifts. After a  best fit is identified for each BCG, we use the K-correction term from the template to compute BCG luminosity, and then the mass-to-light ratio to compute BCG stellar mass. We derive BCG luminosity in DES $z$ band, and BCG stellar mass according to $z$ band luminosity. As an alternative, we also use the \cite{2007AJ....133..734B} K-correction package to derive BCG luminosity, but the result remains unchanged.

\subsection{BCG Luminosity Uncertainty}
\label{sec:lum_err}

We estimate BCG luminosity uncertainty combining BCG magnitude and redshift (see M15) uncertainties. To simplify subsequent analyses, we assume the redshift uncertainty to be statistical (systematic uncertainty is about $\sim$ 0.001, comparing to $\sim$ 0.05 statistical uncertainty, see M15). The redshift uncertainty is taken as 0.001 if archival spectroscopic redshift is available. We ignore K-correction uncertainty as it is well decided.

\subsection{BCG Mass-to-Light Ratio Uncertainty}
\label{sec:m_err}

\begin{figure}
\includegraphics[width=1.0\textwidth]{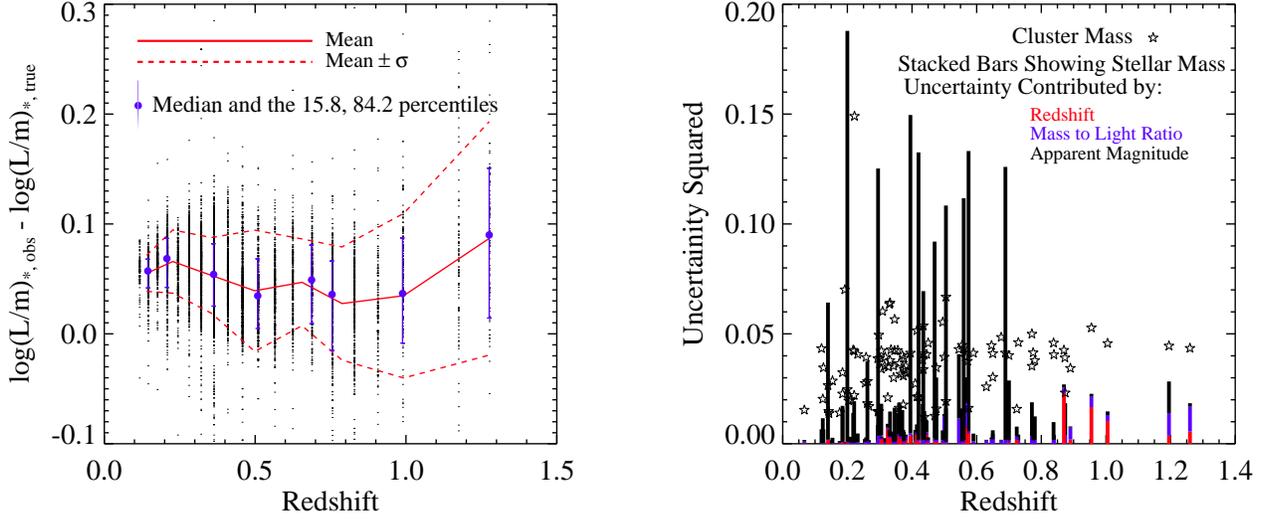}
\caption{(a) We apply our stellar model fitting procedure to the DL07 simulation BCGs using their simulated SDSS {\it g, r, i, z} photometry. Because the fitting procedure and the simulated BCG photometry are based on different SSP models and IMFs, there exists a 0.05 dex systematic offset. The derived mass-to-light ratio also has statistical uncertainty of $\sim$ 0.05 dex at z = 0 and $\sim$ 0.1 dex at z =1. (b) We evaluate the stellar mass uncertainty combining redshift, magnitude and mass-to-light ratio uncertainties. The stellar mass uncertainty is dominated by apparent magnitude uncertainty at z < 0.8, and redshift uncertainty at z > 0.8.}
\label{fig:stellar_comp}
\end{figure}

We estimate BCG stellar mass uncertainty combining BCG luminosity uncertainty and BCG  mass-to-light ratio (MLR) uncertainty. In this section, we pay special attention to estimating the MLR uncertainty from modeling star formation histories, which is the uncertainty from fitting a $\tau$-model to BCGs that formed through merging with galaxies of various star formation histories. We evaluate the uncertainty by applying the stellar population fitting procedure to DL07 BCGs (selected in Section~\ref{sec:sim_select}). We compare the derived BCG MLR to their true values in the simulation.

In the left panel of Figure~\ref{fig:stellar_comp}, we show the difference between the derived and true MLR plotted against redshift. We notice a systematic uncertainty of $\sim$ 0.05 dex, likely caused by the mismatch of IMF in our procedure and in DL07 \citep[a Chabrier IMF produces a mass-to-light ratio 0.05 dex higher than that of a Kroupa IMF, see][]{2011MNRAS.412.1123P}.

We also notice a statistical scatter with the derived values, ranging from 0.05 dex to 0.1 dex with weak dependence on redshift (Figure~\ref{fig:stellar_comp}), but no dependence on cluster mass or BCG stellar mass. We evaluate the uncertainty and covariance for our BCG sample taking the corresponding values in simulation. To elaborate, for each BCG in our sample, we assume its MLR to have been measured 100 times (each BCG is matched to 100 simulation BCGs in Section~\ref{sec:sim_select}), and the error of each measurement is the offset between the derived and true MLR for one simulation BCG. As a result of this set-up, the MLR uncertainty for each BCG contains about 0.05 dex systematic uncertainty and 0.05 to 0.1 dex statistical uncertainty depending on its redshift.

Admittedly, it is more than likely that we are underestimating the BCG MLR uncertainty. In our simulation test, the systematic uncertainty originates from using slightly different SSP models and IMFs (\cite{2009ApJ...699..486C,  2010ApJ...712..833C} SSP models and \cite{2003PASP..115..763C} IMF in our procedure VS the \cite{2003MNRAS.344.1000B} SSP models  and \cite{2001MNRAS.322..231K} IMF in DL07). The statistical uncertainty originates from matching $\tau$ star formation history and fixed metallicity to DL07 BCGs. We have not considered uncertainties associated with SSP models, dust distributions and possible IMF variations

Estimating the uncertainties from these so-called "known unknowns" is difficult. \cite{2009ApJ...699..486C} shows that one may at best recover the MLR of bright red galaxies with 0.15 dex uncertainty at z = 0, or 0.3 dex at z = 2.0. According to this result, we would have under-estimated BCG MLR uncertainty by $\sim$ 0.1 dex. We also experimented with the SSP models from \cite{2005MNRAS.362..799M} and \cite{2003MNRAS.344.1000B}, but the derived MLR differences are lower than 0.1 dex.  

Since the redshift dependence of our estimation is qualitatively similar to that presented in \cite{2009ApJ...699..486C}, it is un-likely that we are affected with our conclusion about BCG redshift evolution. We therefore do not attempt to include additional uncertainties from the "known unknowns". Eventually, the BCG stellar mass uncertainty is dominated by the uncertainty from magnitude measurement or redshift (See Figure~\ref{fig:stellar_comp}b), rather than from MLR.

\section{Covariance and the Likelihood Function}
\label{sec:additional_likelihood}

This appendix provides extra information about the likelihood function presented in Section~\ref{sec:shm}. We assume multivariate normal distribution for $\bf Y$. Combined with a Bernoulli prior distribution for $\bf Q$, the posterior likelihood becomes:

\begin{equation}
\mathcal{L} \propto p({\bf Q})\times|{\bf C}|^{-\frac{1}{2}}exp\big(-\frac{1}{2}{\bf Y}^T  {\bf C }^{-1} {\bf Y}\big).
\end{equation}
Re-write the likelihood at the log scale and ignore the normalization of the probability distribution, we have:
\begin{equation}
  \mathrm{log}~\mathcal{L} =-\frac{1}{2} \mathrm{log} | {\bf C}|  -\frac{1}{2} {\bf Y^{\mathrm{T}} C^{-1} Y } +\mathrm{log}p(\mathrm{\bf Q}).
\end{equation}

Because \begin{equation}
 {\bf Y}={\bf Y}_{\mathrm meas}-{\bf Y}_{\mathrm model},
\end{equation}
The covariance matrix used in the posterior likelihood can be further expanded as :
\begin{equation}
\begin{split}
{\bf C}&= {\bf Cov(Y, Y)} \\
&={\bf Cov(Y_{\mathrm{model}}-Y_{\mathrm{obs}}, \, Y_{\mathrm{model}}-Y_{\mathrm{obs}}) }\\
& ={\bf Cov(Y_{\mathrm{obs}}, Y_{\mathrm{obs}})}+{\bf Cov(Y_{\mathrm{model}}, Y_{\mathrm{model}})}
 \\
&= {\bf Cov(m_{*, obs}, m_{*, obs})+\alpha^2Cov (\mathrm{log}M_{200}, \mathrm{log}M_{200}) }{\bf + \beta^2 Cov(\mathrm{log}(\textnormal{1}+z), \mathrm{log}(\textnormal{1}+z)) } + \sigma ^2 {\bf I}. \\
\end{split}
\end{equation}
We have simplified this expression as Equation~\ref{eq:covariance} in Section~\ref{sec:shm}.

If the covariance matrices for cluster redshift, cluster $M_{200}$ and BCG stellar mass are all diagonal, our posterior likelihood function -- Equation~\ref{eq:likelihood} -- would have the same form as that derived in  \cite{2010arXiv1008.4686H}. In our case, the covariance matrices for $M_{200}$ and BCG stellar masses are not diagonal because of systematic uncertainties (See details in Sections~\ref{sec:mass} to \ref{sec:bcg_lm}).

\section{Table -- BCGs and their host cluster properties}

\end{appendix}
\input{table_bcg.tex}
\end{document}

%% file: author_list.tex
\author{
Y.~Zhang\altaffilmark{1}$^{, \dagger}$,
C.~Miller\altaffilmark{1},
T.~McKay\altaffilmark{1},
P.~Rooney\altaffilmark{2},
A.~E.~Evrard\altaffilmark{1},
A.~K.~Romer\altaffilmark{2},
R.~Perfecto\altaffilmark{3},
J.~Song\altaffilmark{4},
S.~Desai\altaffilmark{5},
J.~Mohr\altaffilmark{5},
H.~Wilcox\altaffilmark{6},
A.~Bermeo\altaffilmark{2},
T.~Jeltema\altaffilmark{7},
D.~Hollowood\altaffilmark{7},
D.~Bacon\altaffilmark{6},
D.~Capozzi\altaffilmark{6},
C.~Collins\altaffilmark{8},
R.~Das\altaffilmark{1},
D.~Gerdes\altaffilmark{1},
C.~Hennig\altaffilmark{5},
M.~Hilton\altaffilmark{9},
B.~Hoyle\altaffilmark{5},
S.~Kay\altaffilmark{10},
A.~Liddle\altaffilmark{11},
R.~G.~Mann\altaffilmark{11},
N.~Mehrtens\altaffilmark{12},
R.~C.~Nichol\altaffilmark{6},
C.~Papovich\altaffilmark{12},
M.~Sahl\'en\altaffilmark{13},
M.~Soares-Santos\altaffilmark{14},
J.~Stott\altaffilmark{15},
P.~T.~Viana\altaffilmark{16,17},
T.~Abbott\altaffilmark{18},
F.~B.~Abdalla\altaffilmark{19},
M.~Banerji\altaffilmark{20,21},
A.H.~Bauer\altaffilmark{22},
A.~Benoit-L{\'e}vy\altaffilmark{19},
E.~Bertin\altaffilmark{23},
D.~Brooks\altaffilmark{19},
E.~Buckley-Geer\altaffilmark{14},
D.~L.~Burke\altaffilmark{24,25},
A.~Carnero~Rosell\altaffilmark{26,27},
F.~J.~Castander\altaffilmark{22},
H.~T.~Diehl\altaffilmark{14},
P.~Doel\altaffilmark{19},
C.~E Cunha\altaffilmark{24},
T.~F.~Eifler\altaffilmark{28,29},
A.~Fausti Neto\altaffilmark{26},
E.~Fernandez\altaffilmark{30},
B.~Flaugher\altaffilmark{14},
P.~Fosalba\altaffilmark{22},
J.~Frieman\altaffilmark{14,31},
E.~Gaztanaga\altaffilmark{22},
D.~Gruen\altaffilmark{32,33},
R.~A.~Gruendl\altaffilmark{34,35},
K.~Honscheid\altaffilmark{36,37},
D.~James\altaffilmark{18},
K.~Kuehn\altaffilmark{38},
N.~Kuropatkin\altaffilmark{14},
O.~Lahav\altaffilmark{19},
M.~A.~G.~Maia\altaffilmark{26,27},
M.~Makler\altaffilmark{39},
J.~L.~Marshall\altaffilmark{12},
Paul~Martini\altaffilmark{36,40},
R.~Miquel\altaffilmark{30},
R.~Ogando\altaffilmark{26,27},
A.~A.~Plazas\altaffilmark{41,29},
A.~Roodman\altaffilmark{24,25},
E.~S.~Rykoff\altaffilmark{24,25},
M.~Sako\altaffilmark{28},
E.~Sanchez\altaffilmark{42},
V.~Scarpine\altaffilmark{14},
M.~Schubnell\altaffilmark{1},
I.~Sevilla\altaffilmark{42,34},
R.~C.~Smith\altaffilmark{18},
F.~Sobreira\altaffilmark{14,26},
E.~Suchyta\altaffilmark{36,37},
M.~E.~C.~Swanson\altaffilmark{35},
G.~Tarle\altaffilmark{1},
J.~Thaler\altaffilmark{43},
D.~Tucker\altaffilmark{14},
V.~Vikram\altaffilmark{44},
L.~N.~da Costa\altaffilmark{26,27}
}
 
\altaffiltext{1}{Department of Physics, University of Michigan, Ann Arbor, MI 48109, USA}
\altaffiltext{2}{Department of Physics and Astronomy, Pevensey Building, University of Sussex, Brighton, BN1 9QH, UK}
\altaffiltext{3}{Department of Astronomy, Yale University, New Haven, CT, 06511-8903}
\altaffiltext{4}{Korea Astronomy and Space Science Institute, 776, Daedeokdae-ro,Yuseong-gu, Daejeon, 305-348, Republic of Korea}
\altaffiltext{5}{Department of Physics, Ludwig-Maximilians-Universit\"at, Scheinerstr.\ 1, 81679 M\"unchen, Germany}
\altaffiltext{6}{Institute of Cosmology \& Gravitation, University of Portsmouth, Portsmouth, PO1 3FX, UK}
\altaffiltext{7}{Department of Physics and Santa Cruz Institute for Particle Physics, University of California, Santa Cruz, CA 95064, USA}
\altaffiltext{8}{Astrophysics Research Institute, Liverpool John Moores University, IC2, Liverpool Science Park, 146 Brownlow Hill, Liverpool L3 5RF, UK}
\altaffiltext{9}{Astrophysics and Cosmology Research Unit, School of Mathematics, Statistics and Computer Science, University of KwaZulu\-Natal, Westville Campus, Durban 4000, South Africa}
\altaffiltext{10}{Jodrell Bank Centre for Astrophysics, School of Physics and Astronomy, The University of Manchester, Manchester M13 9PL, UK}
\altaffiltext{11}{Institute for Astronomy, University of Edinburgh, Royal Observatory, Blackford Hill, Edinburgh EH9 3HJ, United Kingdom}
\altaffiltext{12}{George P. and Cynthia Woods Mitchell Institute for Fundamental Physics and Astronomy, and Department of Physics and Astronomy, Texas A\&M University, College Station, TX 77843,  USA}
\altaffiltext{13}{BIPAC, Department of Physics, University of Oxford,Denys Wilkinson Building, 1 Keble Road, Oxford OX1 3RH, UK}
\altaffiltext{14}{Fermi National Accelerator Laboratory, P. O. Box 500, Batavia, IL 60510, USA}
\altaffiltext{15}{Sub-department of Astrophysics, Department of Physics, University of Oxford, Denys Wilkinson Building, Keble Road, Oxford OX1 3RH, UK}
\altaffiltext{16}{Instituto de Astrof\'{\i}sica e Ci\^{e}ncias do Espa\c{c}o, Universidade do Porto, CAUP, Rua das Estrelas, 4150-762 Porto, Portugal}
\altaffiltext{17}{Departamento de F\'{\i}sica e Astronomia, Faculdade de Ci\^{e}ncias, Universidade do Porto, Rua do Campo Alegre 687, 4169-007 Porto, Portugal}
\altaffiltext{18}{Cerro Tololo Inter-American Observatory, National Optical Astronomy Observatory, Casilla 603, La Serena, Chile}
\altaffiltext{19}{Department of Physics \& Astronomy, University College London, Gower Street, London, WC1E 6BT, UK}
\altaffiltext{20}{Kavli Institute for Cosmology, University of Cambridge, Madingley Road, Cambridge CB3 0HA, UK}
\altaffiltext{21}{Institute of Astronomy, University of Cambridge, Madingley Road, Cambridge CB3 0HA, UK}
\altaffiltext{22}{Institut de Ci\`encies de l'Espai, IEEC-CSIC, Campus UAB, Facultat de Ci\`encies, Torre C5 par-2, 08193 Bellaterra, Barcelona, Spain}
\altaffiltext{23}{Institut d'Astrophysique de Paris, Univ. Pierre et Marie Curie \& CNRS UMR7095, F-75014 Paris, France}
\altaffiltext{24}{Kavli Institute for Particle Astrophysics \& Cosmology, P. O. Box 2450, Stanford University, Stanford, CA 94305, USA}
\altaffiltext{25}{SLAC National Accelerator Laboratory, Menlo Park, CA 94025, USA}
\altaffiltext{26}{Laborat\'orio Interinstitucional de e-Astronomia - LIneA, Rua Gal. Jos\'e Cristino 77, Rio de Janeiro, RJ - 20921-400, Brazil}
\altaffiltext{27}{Observat\'orio Nacional, Rua Gal. Jos\'e Cristino 77, Rio de Janeiro, RJ - 20921-400, Brazil}
\altaffiltext{28}{Department of Physics and Astronomy, University of Pennsylvania, Philadelphia, PA 19104, USA}
\altaffiltext{29}{Jet Propulsion Laboratory, California Institute of Technology, 4800 Oak Grove Dr., Pasadena, CA 91109, USA}
\altaffiltext{30}{Institut de F\'{\i}sica d'Altes Energies, Universitat Aut\`onoma de Barcelona, E-08193 Bellaterra, Barcelona, Spain}
\altaffiltext{31}{Kavli Institute for Cosmological Physics, University of Chicago, Chicago, IL 60637, USA}
\altaffiltext{32}{Max Planck Institute for Extraterrestrial Physics, Giessenbachstrasse, 85748 Garching, Germany}
\altaffiltext{33}{University Observatory Munich, Scheinerstrasse 1, 81679 Munich, Germany}
\altaffiltext{34}{Department of Astronomy, University of Illinois,1002 W. Green Street, Urbana, IL 61801, USA}
\altaffiltext{35}{National Center for Supercomputing Applications, 1205 West Clark St., Urbana, IL 61801, USA}
\altaffiltext{36}{Center for Cosmology and Astro-Particle Physics, The Ohio State University, Columbus, OH 43210, USA}
\altaffiltext{37}{Department of Physics, The Ohio State University, Columbus, OH 43210, USA}
\altaffiltext{38}{Australian Astronomical Observatory, North Ryde, NSW 2113, Australia}
\altaffiltext{39}{ICRA, Centro Brasileiro de Pesquisas F\'isicas, Rua Dr. Xavier Sigaud 150, CEP 22290-180, Rio de Janeiro, RJ, Brazil}
\altaffiltext{40}{Department of Astronomy, The Ohio State University, Columbus, OH 43210, USA}
\altaffiltext{41}{Brookhaven National Laboratory, Bldg 510, Upton, NY 11973, USA}
\altaffiltext{42}{Centro de Investigaciones Energ\'eticas, Medioambientales y Tecnol\'ogicas (CIEMAT), Madrid, Spain}
\altaffiltext{43}{Department of Physics, University of Illinois, 1110 W. Green St., Urbana, IL 61801, USA}
\altaffiltext{44}{Argonne National Laboratory, 9700 South Cass Avenue, Lemont, IL 60439, USA}

%% file: table_bcg.tex
\LongTables
\begin{deluxetable*}{cccccccc}
\tablecaption{ \label{tbl:bcgs}}
\tabletypesize{\footnotesize}
\startdata
& Cluster Name  & $M_{200}$ & Cluster Redshift & BCG Stellar Mass & BCG DES $z$ Luminosity \\
      1 & XMMXCSJ003545.5-431758.6  & 14.17 $\pm$  0.20 & 0.650 $\pm$ 0.002 & 11.48 $\pm$  0.07 & -23.45 $\pm$  0.05 \\
      2 & XMMXCSJ003548.1-432232.8  & 14.30 $\pm$  0.16 & 0.630 $\pm$ 0.002 & 11.53 $\pm$  0.06 & -23.78 $\pm$  0.04 \\
      3 & XMMXCSJ021442.21-043315.3 & 14.72 $\pm$  0.12 & 0.142 $\pm$ 0.001 & 11.75 $\pm$  0.06 & -24.13 $\pm$  0.05 \\
      4 & XMMXCSJ021547.28-045031.4 & 13.75 $\pm$  0.23 & 0.955 $\pm$ 0.108 & 11.75 $\pm$  0.15 & -24.16 $\pm$  0.32 \\
      5 & XMMXCSJ021611.69-041422.8 & 14.44 $\pm$  0.17 & 0.154 $\pm$ 0.001 & 11.52 $\pm$  0.07 & -23.77 $\pm$  0.12 \\
      6 & XMMXCSJ021735.7-051315.8  & 14.25 $\pm$  0.21 & 0.646 $\pm$ 0.002 & 11.34 $\pm$  0.06 & -23.38 $\pm$  0.07 \\
      7 & XMMXCSJ021807.88-054557.3 & 14.56 $\pm$  0.20 & 0.689 $\pm$ 0.002 & 11.29 $\pm$  0.39 & -23.07 $\pm$  0.96 \\
      8 & XMMXCSJ021826.0-045942.7  & 14.26 $\pm$  0.20 & 0.873 $\pm$ 0.002 & 10.20 $\pm$  0.13 & -21.10 $\pm$  0.30 \\
      9 & XMMXCSJ021833.0-050101.5  & 14.19 $\pm$  0.15 & 0.874 $\pm$ 0.002 & 11.37 $\pm$  0.14 & -23.70 $\pm$  0.30 \\
     10 & XMMXCSJ021837.90-054037.0 & 13.92 $\pm$  0.20 & 0.317 $\pm$ 0.001 & 11.65 $\pm$  0.08 & -24.10 $\pm$  0.11 \\
     11 & XMMXCSJ021843.7-053253.3  & 13.94 $\pm$  0.18 & 0.381 $\pm$ 0.001 & 11.10 $\pm$  0.09 & -22.76 $\pm$  0.16 \\
     12 & XMMXCSJ021914.5-045053.2  & 13.83 $\pm$  0.39 & 0.222 $\pm$ 0.001 & 11.40 $\pm$  0.08 & -23.13 $\pm$  0.08 \\
     13 & XMMXCSJ021940.78-055043.7 & 14.23 $\pm$  0.21 & 0.729 $\pm$ 0.002 & 11.30 $\pm$  0.09 & -23.74 $\pm$  0.14 \\
     14 & XMMXCSJ021944.3-045326.8  & 14.24 $\pm$  0.25 & 0.332 $\pm$ 0.001 & 10.96 $\pm$  0.11 & -22.20 $\pm$  0.23 \\
     15 & XMMXCSJ022154.83-054519.0 & 14.51 $\pm$  0.20 & 0.258 $\pm$ 0.001 & 11.67 $\pm$  0.09 & -23.63 $\pm$  0.11 \\
     16 & XMMXCSJ022204.5-043246.3  & 14.26 $\pm$  0.19 & 0.315 $\pm$ 0.001 & 11.23 $\pm$  0.08 & -23.08 $\pm$  0.12 \\
     17 & XMMXCSJ022233.02-045803.5 & 14.41 $\pm$  0.20 & 0.839 $\pm$ 0.002 & 11.68 $\pm$  0.07 & -24.14 $\pm$  0.12 \\
     18 & XMMXCSJ022302.5-043619.6  & 13.84 $\pm$  0.21 & 1.260 $\pm$ 0.085 & 11.79 $\pm$  0.16 & -24.01 $\pm$  0.20 \\
     19 & XMMXCSJ022318.3-051209.8  & 14.00 $\pm$  0.19 & 0.332 $\pm$ 0.001 & 11.15 $\pm$  0.08 & -23.17 $\pm$  0.03 \\
     20 & XMMXCSJ022318.37-052707.6 & 13.69 $\pm$  0.14 & 0.211 $\pm$ 0.001 & 11.50 $\pm$  0.09 & -23.47 $\pm$  0.12 \\
     21 & XMMXCSJ022347.50-025134.4 & 13.86 $\pm$  0.15 & 0.185 $\pm$ 0.007 & 11.52 $\pm$  0.14 & -23.64 $\pm$  0.31 \\
     22 & XMMXCSJ022350.88-053643.9 & 14.85 $\pm$  0.20 & 0.500 $\pm$ 0.002 & 11.47 $\pm$  0.12 & -23.74 $\pm$  0.08 \\
     23 & XMMXCSJ022356.32-030556.8 & 14.04 $\pm$  0.20 & 0.300 $\pm$ 0.010 & 11.31 $\pm$  0.10 & -23.18 $\pm$  0.20 \\
     24 & XMMXCSJ022357.6-043519.7  & 14.44 $\pm$  0.19 & 0.572 $\pm$ 0.002 & 11.24 $\pm$  0.11 & -23.54 $\pm$  0.20 \\
     25 & XMMXCSJ022401.1-050542.2  & 14.04 $\pm$  0.19 & 0.327 $\pm$ 0.001 & 11.43 $\pm$  0.08 & -23.42 $\pm$  0.12 \\
     26 & XMMXCSJ022433.9-041442.7  & 14.04 $\pm$  0.14 & 0.262 $\pm$ 0.001 & 11.52 $\pm$  0.21 & -23.56 $\pm$  0.49 \\
     27 & XMMXCSJ022512.2-062307.9  & 14.41 $\pm$  0.14 & 0.204 $\pm$ 0.001 & 11.74 $\pm$  0.08 & -24.18 $\pm$  0.09 \\
     28 & XMMXCSJ022512.31-053112.3 & 13.62 $\pm$  0.20 & 0.294 $\pm$ 0.001 & 11.54 $\pm$  0.10 & -23.45 $\pm$  0.14 \\
     29 & XMMXCSJ022524.7-044043.9  & 14.18 $\pm$  0.13 & 0.267 $\pm$ 0.001 & 11.60 $\pm$  0.07 & -23.65 $\pm$  0.04 \\
     30 & XMMXCSJ022529.8-041432.7  & 14.00 $\pm$  0.12 & 0.143 $\pm$ 0.001 & 11.27 $\pm$  0.07 & -23.14 $\pm$  0.07 \\
     31 & XMMXCSJ022532.42-035502.4 & 14.02 $\pm$  0.22 & 0.771 $\pm$ 0.002 & 11.52 $\pm$  0.08 & -23.75 $\pm$  0.09 \\
     32 & XMMXCSJ022549.02-055339.3 & 14.42 $\pm$  0.20 & 0.232 $\pm$ 0.001 & 11.58 $\pm$  0.09 & -23.55 $\pm$  0.16 \\
     33 & XMMXCSJ022559.68-024932.4 & 14.04 $\pm$  0.21 & 0.305 $\pm$ 0.017 & 11.18 $\pm$  0.15 & -22.91 $\pm$  0.33 \\
     34 & XMMXCSJ022632.5-054651.9  & 14.40 $\pm$  0.20 & 0.445 $\pm$ 0.026 & 11.54 $\pm$  0.10 & -23.86 $\pm$  0.21 \\
     35 & XMMXCSJ022634.75-040408.0 & 14.22 $\pm$  0.21 & 0.346 $\pm$ 0.001 & 11.67 $\pm$  0.07 & -24.13 $\pm$  0.04 \\
     36 & XMMXCSJ022722.82-032147.3 & 14.38 $\pm$  0.13 & 0.360 $\pm$ 0.016 & 11.57 $\pm$  0.14 & -23.81 $\pm$  0.32 \\
     37 & XMMXCSJ022726.4-043206.8  & 14.03 $\pm$  0.25 & 0.309 $\pm$ 0.001 & 11.58 $\pm$  0.07 & -23.83 $\pm$  0.03 \\
     38 & XMMXCSJ022740.4-045129.9  & 14.09 $\pm$  0.12 & 0.295 $\pm$ 0.001 & 11.37 $\pm$  0.06 & -23.66 $\pm$  0.05 \\
     39 & XMMXCSJ022802.85-045101.1 & 14.57 $\pm$  0.22 & 0.297 $\pm$ 0.001 & 11.53 $\pm$  0.09 & -23.79 $\pm$  0.06 \\
     40 & XMMXCSJ022808.41-053553.2 & 13.98 $\pm$  0.26 & 0.192 $\pm$ 0.001 & 11.50 $\pm$  0.08 & -23.53 $\pm$  0.06 \\
     41 & XMMXCSJ022827.3-042542.5  & 14.37 $\pm$  0.20 & 0.433 $\pm$ 0.001 & 10.46 $\pm$  0.15 & -21.66 $\pm$  0.36 \\
     42 & XMMXCSJ022829.83-031257.2 & 14.14 $\pm$  0.20 & 0.370 $\pm$ 0.000 & 11.63 $\pm$  0.13 & -23.98 $\pm$  0.30 \\
     43 & XMMXCSJ023052.5-045128.7  & 14.41 $\pm$  0.20 & 0.590 $\pm$ 0.002 & 11.34 $\pm$  0.10 & -25.00 $\pm$  0.15 \\
     44 & XMMXCSJ033151.23-274936.2 & 13.77 $\pm$  0.22 & 0.676 $\pm$ 0.002 & 11.19 $\pm$  0.06 & -23.29 $\pm$  0.04 \\
     45 & XMMXCSJ034004.0-283150.6  & 14.02 $\pm$  0.17 & 0.262 $\pm$ 0.001 & 11.57 $\pm$  0.06 & -23.47 $\pm$  0.05 \\
     46 & XMMXCSJ034025.95-284025.4 & 14.46 $\pm$  0.12 & 0.067 $\pm$ 0.001 & 12.27 $\pm$  0.06 & -22.38 $\pm$  0.06 \\
     47 & XMMXCSJ034026.03-282835.8 & 14.44 $\pm$  0.20 & 0.375 $\pm$ 0.007 & 11.58 $\pm$  0.07 & -23.93 $\pm$  0.06 \\
     48 & XMMXCSJ041646.0-552510.4  & 14.14 $\pm$  0.17 & 0.410 $\pm$ 0.008 & 11.61 $\pm$  0.10 & -24.08 $\pm$  0.18 \\
     49 & XMMXCSJ042017.8-503155.0  & 14.24 $\pm$  0.11 & 0.470 $\pm$ 0.015 & 11.60 $\pm$  0.23 & -24.03 $\pm$  0.56 \\
     50 & XMMXCSJ042226.36-514029.7 & 14.35 $\pm$  0.13 & 0.575 $\pm$ 0.039 & 11.69 $\pm$  0.34 & -24.29 $\pm$  0.85 \\
     51 & XMMXCSJ043218.04-610356.5 & 14.50 $\pm$  0.23 & 0.435 $\pm$ 0.021 & 11.67 $\pm$  0.09 & -24.18 $\pm$  0.19 \\
     52 & XMMXCSJ043318.93-614013.7 & 14.55 $\pm$  0.21 & 0.545 $\pm$ 0.007 & 11.11 $\pm$  0.24 & -22.98 $\pm$  0.51 \\
     53 & XMMXCSJ043706.81-541413.0 & 14.33 $\pm$  0.14 & 0.505 $\pm$ 0.006 & 11.42 $\pm$  0.20 & -23.56 $\pm$  0.47 \\
     54 & XMMXCSJ043708.09-542908.8 & 14.57 $\pm$  0.20 & 0.565 $\pm$ 0.041 & 11.14 $\pm$  0.16 & -22.91 $\pm$  0.28 \\
     55 & XMMXCSJ043818.09-541917.5 & 15.05 $\pm$  0.13 & 0.475 $\pm$ 0.005 & 11.84 $\pm$  0.24 & -24.63 $\pm$  0.58 \\
     56 & XMMXCSJ043940.82-542412.9 & 14.67 $\pm$  0.20 & 0.700 $\pm$ 0.015 & 11.82 $\pm$  0.20 & -24.47 $\pm$  0.47 \\
     57 & XMMXCSJ045421.2-531531.1  & 13.91 $\pm$  0.14 & 0.325 $\pm$ 0.028 & 11.15 $\pm$  0.11 & -22.49 $\pm$  0.23 \\
     58 & XMMXCSJ045506.1-532343.2  & 14.40 $\pm$  0.18 & 0.435 $\pm$ 0.019 & 11.51 $\pm$  0.14 & -23.72 $\pm$  0.34 \\
     59 & XMMXCSJ051141.31-515420.8 & 14.44 $\pm$  0.13 & 0.724 $\pm$ 0.028 & 11.57 $\pm$  0.10 & -24.12 $\pm$  0.22 \\
     60 & XMMXCSJ051636.81-543113.3 & 14.84 $\pm$  0.12 & 0.295 $\pm$ 0.001 & 12.15 $\pm$  0.37 & -24.69 $\pm$  0.92 \\
     61 & XMMXCSJ065829.1-555641.5  & 15.12 $\pm$  0.12 & 0.297 $\pm$ 0.001 & 11.77 $\pm$  0.16 & -24.10 $\pm$  0.33 \\
     62 & XMMXCSJ065860.0-560926.8  & 14.23 $\pm$  0.21 & 0.335 $\pm$ 0.015 & 11.69 $\pm$  0.08 & -24.13 $\pm$  0.12 \\
     63 & XMMXCSJ095737.12+023426.0 & 14.75 $\pm$  0.18 & 0.373 $\pm$ 0.001 & 11.99 $\pm$  0.08 & -24.57 $\pm$  0.13 \\
     64 & XMMXCSJ095823.49+024920.2 & 14.25 $\pm$  0.19 & 0.343 $\pm$ 0.001 & 11.56 $\pm$  0.08 & -23.64 $\pm$  0.13 \\
     65 & XMMXCSJ095846.90+021550.8 & 14.39 $\pm$  0.19 & 0.771 $\pm$ 0.002 & 11.38 $\pm$  0.13 & -23.62 $\pm$  0.27 \\
     66 & XMMXCSJ095924.73+014615.7 & 13.86 $\pm$  0.14 & 0.124 $\pm$ 0.001 & 11.22 $\pm$  0.10 & -23.02 $\pm$  0.20 \\
     67 & XMMXCSJ095931.63+022657.2 & 14.74 $\pm$  0.21 & 0.356 $\pm$ 0.001 & 11.57 $\pm$  0.13 & -23.70 $\pm$  0.30 \\
     68 & XMMXCSJ095944.68+023619.8 & 13.53 $\pm$  0.17 & 0.343 $\pm$ 0.001 & 11.62 $\pm$  0.10 & -24.12 $\pm$  0.19 \\
     69 & XMMXCSJ095947.18+025529.1 & 14.29 $\pm$  0.19 & 0.126 $\pm$ 0.001 & 10.95 $\pm$  0.10 & -22.16 $\pm$  0.18 \\
     70 & XMMXCSJ095951.46+014051.8 & 14.25 $\pm$  0.13 & 0.373 $\pm$ 0.001 & 11.69 $\pm$  0.09 & -24.08 $\pm$  0.17 \\
     71 & XMMXCSJ100021.72+022329.3 & 13.83 $\pm$  0.15 & 0.221 $\pm$ 0.001 & 11.40 $\pm$  0.10 & -23.49 $\pm$  0.18 \\
     72 & XMMXCSJ100027.26+022135.9 & 13.98 $\pm$  0.21 & 0.221 $\pm$ 0.001 & 11.33 $\pm$  0.15 & -23.17 $\pm$  0.32 \\
     73 & XMMXCSJ100043.28+014608.0 & 14.14 $\pm$  0.24 & 0.346 $\pm$ 0.001 & 11.58 $\pm$  0.13 & -23.74 $\pm$  0.28 \\
     74 & XMMXCSJ100047.16+015917.0 & 14.37 $\pm$  0.11 & 0.438 $\pm$ 0.001 & 11.45 $\pm$  0.14 & -23.62 $\pm$  0.32 \\
     75 & XMMXCSJ100109.18+013336.0 & 13.93 $\pm$  0.20 & 0.435 $\pm$ 0.002 & 11.64 $\pm$  0.13 & -24.19 $\pm$  0.29 \\
     76 & XMMXCSJ100141.74+022538.0 & 14.32 $\pm$  0.21 & 0.120 $\pm$ 0.001 & 11.58 $\pm$  0.09 & -23.44 $\pm$  0.17 \\
     77 & XMMXCSJ100142.56+014059.4 & 14.16 $\pm$  0.21 & 0.218 $\pm$ 0.001 & 11.25 $\pm$  0.12 & -23.24 $\pm$  0.24 \\
     78 & XMMXCSJ100201.42+021334.2 & 14.96 $\pm$  0.21 & 0.838 $\pm$ 0.002 & 11.57 $\pm$  0.10 & -23.95 $\pm$  0.22 \\
     79 & XMMXCSJ232737.63-541610.0 & 14.56 $\pm$  0.21 & 1.005 $\pm$ 0.096 & 11.43 $\pm$  0.14 & -24.08 $\pm$  0.29 \\
     80 & XMMXCSJ232810.21-555015.8 & 14.79 $\pm$  0.19 & 0.890 $\pm$ 0.035 & 11.62 $\pm$  0.09 & -24.07 $\pm$  0.13 \\
     81 & XMMXCSJ232900.4-533931.7  & 13.86 $\pm$  0.17 & 0.255 $\pm$ 0.004 & 11.68 $\pm$  0.08 & -23.95 $\pm$  0.09 \\
     82 & XMMXCSJ233000.57-543706.4 & 14.21 $\pm$  0.12 & 0.176 $\pm$ 0.001 & 11.67 $\pm$  0.07 & -23.99 $\pm$  0.06 \\
     83 & XMMXCSJ233003.40-541415.6 & 13.94 $\pm$  0.23 & 0.415 $\pm$ 0.011 & 11.42 $\pm$  0.07 & -23.55 $\pm$  0.09 \\
     84 & XMMXCSJ233037.38-554340.2 & 14.09 $\pm$  0.25 & 0.330 $\pm$ 0.017 & 11.58 $\pm$  0.10 & -23.86 $\pm$  0.18 \\
     85 & XMMXCSJ233135.2-562754.1  & 14.42 $\pm$  0.18 & 0.185 $\pm$ 0.005 & 11.48 $\pm$  0.18 & -23.53 $\pm$  0.41 \\
     86 & XMMXCSJ233204.99-551242.8 & 13.82 $\pm$  0.18 & 0.305 $\pm$ 0.014 & 11.39 $\pm$  0.09 & -23.43 $\pm$  0.14 \\
     87 & XMMXCSJ233215.97-544205.3 & 14.21 $\pm$  0.19 & 0.360 $\pm$ 0.022 & 11.74 $\pm$  0.09 & -24.32 $\pm$  0.18 \\
     88 & XMMXCSJ233331.89-554753.4 & 14.53 $\pm$  0.21 & 1.195 $\pm$ 0.065 & 11.78 $\pm$  0.18 & -23.61 $\pm$  0.33 \\
     89 & XMMXCSJ233346.00-553826.9 & 14.53 $\pm$  0.19 & 0.780 $\pm$ 0.000 & 11.60 $\pm$  0.12 & -23.97 $\pm$  0.26 \\
     90 & XMMXCSJ233406.45-554710.8 & 14.46 $\pm$  0.20 & 0.775 $\pm$ 0.000 & 11.49 $\pm$  0.06 & -23.74 $\pm$  0.12 \\
     91 & XMMXCSJ233421.4-541054.6  & 14.22 $\pm$  0.18 & 0.365 $\pm$ 0.017 & 11.57 $\pm$  0.10 & -23.82 $\pm$  0.19 \\
     92 & XMMXCSJ233429.10-543618.3 & 14.42 $\pm$  0.21 & 0.450 $\pm$ 0.009 & 11.64 $\pm$  0.06 & -24.20 $\pm$  0.10 \\
     93 & XMMXCSJ233528.43-543501.1 & 14.63 $\pm$  0.21 & 0.870 $\pm$ 0.118 & 11.60 $\pm$  0.16 & -24.13 $\pm$  0.37 \\
     94 & XMMXCSJ233637.1-524408.2  & 13.90 $\pm$  0.21 & 0.560 $\pm$ 0.011 & 11.65 $\pm$  0.37 & -24.17 $\pm$  0.93 \\
     95 & XMMXCSJ233706.8-541911.5  & 14.23 $\pm$  0.26 & 0.505 $\pm$ 0.008 & 11.57 $\pm$  0.27 & -24.01 $\pm$  0.67 \\
     96 & XMMXCSJ233745.46-562758.5 & 14.27 $\pm$  0.19 & 0.380 $\pm$ 0.018 & 11.36 $\pm$  0.07 & -23.37 $\pm$  0.11 \\
     97 & XMMXCSJ233836.3-543740.3  & 14.52 $\pm$  0.18 & 0.375 $\pm$ 0.006 & 11.67 $\pm$  0.08 & -24.18 $\pm$  0.10 \\
     98 & XMMXCSJ234119.2-560400.2  & 14.39 $\pm$  0.19 & 0.475 $\pm$ 0.014 & 11.52 $\pm$  0.28 & -23.77 $\pm$  0.70 \\
     99 & XMMXCSJ234142.5-555747.7  & 14.37 $\pm$  0.16 & 0.200 $\pm$ 0.005 & 11.80 $\pm$  0.44 & -24.33 $\pm$  1.09 \\
    100 & XMMXCSJ234145.4-564000.7  & 14.23 $\pm$  0.24 & 0.495 $\pm$ 0.009 & 11.45 $\pm$  0.09 & -23.77 $\pm$  0.18 \\
    101 & XMMXCSJ234201.2-553253.8  & 14.27 $\pm$  0.21 & 0.555 $\pm$ 0.005 & 11.26 $\pm$  0.26 & -23.26 $\pm$  0.64 \\
    102 & XMMXCSJ234231.5-562106.8  & 14.30 $\pm$  0.15 & 0.415 $\pm$ 0.027 & 11.53 $\pm$  0.11 & -23.80 $\pm$  0.22 \\
    103 & XMMXCSJ234448.2-561728.2  & 14.30 $\pm$  0.17 & 0.650 $\pm$ 0.006 & 11.61 $\pm$  0.25 & -23.86 $\pm$  0.59 \\
    104 & XMMXCSJ234730.8-553320.5  & 14.05 $\pm$  0.20 & 0.395 $\pm$ 0.025 & 11.72 $\pm$  0.40 & -24.16 $\pm$  1.00 \\
    105 & XMMXCSJ235009.5-551957.9  & 14.08 $\pm$  0.15 & 0.420 $\pm$ 0.015 & 11.60 $\pm$  0.35 & -24.08 $\pm$  0.85 \\
    106 & XMMXCSJ235059.5-552206.1  & 14.20 $\pm$  0.16 & 0.140 $\pm$ 0.006 & 11.35 $\pm$  0.26 & -23.28 $\pm$  0.63 \\
\enddata
\end{deluxetable*}
\clearpage

%% file: bcg_evolution.bbl
\begin{thebibliography}{143}
\expandafter\ifx\csname natexlab\endcsname\relax\def\natexlab#1{#1}\fi

\bibitem[{{Andreon}(2002)}]{2002A&A...382..495A}
{Andreon}, S. 2002, \aap, 382, 495

\bibitem[{{Bai} {et~al.}(2014){Bai}, {Yee}, {Yan}, {Lee}, {Gilbank},
  {Ellingson}, {Barrientos}, {Gladders}, {Hsieh}, \&
  {Li}}]{2014ApJ...789..134B}
{Bai}, L., {Yee}, H.~K.~C., {Yan}, R., {et~al.} 2014, \apj, 789, 134

\bibitem[{{Barden} {et~al.}(2012){Barden}, {H{\"a}u{\ss}ler}, {Peng},
  {McIntosh}, \& {Guo}}]{2012MNRAS.422..449B}
{Barden}, M., {H{\"a}u{\ss}ler}, B., {Peng}, C.~Y., {McIntosh}, D.~H., \&
  {Guo}, Y. 2012, \mnras, 422, 449

\bibitem[{{Behroozi} {et~al.}(2013){Behroozi}, {Wechsler}, \&
  {Conroy}}]{2013ApJ...770...57B}
{Behroozi}, P.~S., {Wechsler}, R.~H., \& {Conroy}, C. 2013, \apj, 770, 57

\bibitem[{{Berg{\'e}} {et~al.}(2013){Berg{\'e}}, {Gamper}, {R{\'e}fr{\'e}gier},
  \& {Amara}}]{2013A&C.....1...23B}
{Berg{\'e}}, J., {Gamper}, L., {R{\'e}fr{\'e}gier}, A., \& {Amara}, A. 2013,
  Astronomy and Computing, 1, 23

\bibitem[{{Bernardi} {et~al.}(2007){Bernardi}, {Hyde}, {Sheth}, {Miller}, \&
  {Nichol}}]{2007AJ....133.1741B}
{Bernardi}, M., {Hyde}, J.~B., {Sheth}, R.~K., {Miller}, C.~J., \& {Nichol},
  R.~C. 2007, \aj, 133, 1741

\bibitem[{{Bernardi} {et~al.}(2013){Bernardi}, {Meert}, {Sheth}, {Vikram},
  {Huertas-Company}, {Mei}, \& {Shankar}}]{2013MNRAS.436..697B}
{Bernardi}, M., {Meert}, A., {Sheth}, R.~K., {et~al.} 2013, \mnras, 436, 697

\bibitem[{{Bernardi} {et~al.}(2014){Bernardi}, {Meert}, {Vikram},
  {Huertas-Company}, {Mei}, {Shankar}, \& {Sheth}}]{2014MNRAS.443..874B}
{Bernardi}, M., {Meert}, A., {Vikram}, V., {et~al.} 2014, \mnras, 443, 874

\bibitem[{{Bernstein} {et~al.}(2002){Bernstein}, {Freedman}, \&
  {Madore}}]{2002ApJ...571..107B}
{Bernstein}, R.~A., {Freedman}, W.~L., \& {Madore}, B.~F. 2002, \apj, 571, 107

\bibitem[{{Bertin}(2011)}]{2011ASPC..442..435B}
{Bertin}, E. 2011, in Astronomical Society of the Pacific Conference Series,
  Vol. 442, Astronomical Data Analysis Software and Systems XX, ed. I.~N.
  {Evans}, A.~{Accomazzi}, D.~J. {Mink}, \& A.~H. {Rots}, 435

\bibitem[{{Bertin} \& {Arnouts}(1996)}]{1996A&AS..117..393B}
{Bertin}, E., \& {Arnouts}, S. 1996, \aaps, 117, 393

\bibitem[{{Blanton} \& {Roweis}(2007)}]{2007AJ....133..734B}
{Blanton}, M.~R., \& {Roweis}, S. 2007, \aj, 133, 734

\bibitem[{{Blanton} {et~al.}(2001){Blanton}, {Dalcanton}, {Eisenstein},
  {Loveday}, {Strauss}, {SubbaRao}, {Weinberg}, {Anderson}, {Annis}, {Bahcall},
  {Bernardi}, {Brinkmann}, {Brunner}, {Burles}, {Carey}, {Castander},
  {Connolly}, {Csabai}, {Doi}, {Finkbeiner}, {Friedman}, {Frieman}, {Fukugita},
  {Gunn}, {Hennessy}, {Hindsley}, {Hogg}, {Ichikawa}, {Ivezi{\'c}}, {Kent},
  {Knapp}, {Lamb}, {Leger}, {Long}, {Lupton}, {McKay}, {Meiksin}, {Merelli},
  {Munn}, {Narayanan}, {Newcomb}, {Nichol}, {Okamura}, {Owen}, {Pier}, {Pope},
  {Postman}, {Quinn}, {Rockosi}, {Schlegel}, {Schneider}, {Shimasaku},
  {Siegmund}, {Smee}, {Snir}, {Stoughton}, {Stubbs}, {Szalay}, {Szokoly},
  {Thakar}, {Tremonti}, {Tucker}, {Uomoto}, {Vanden Berk}, {Vogeley},
  {Waddell}, {Yanny}, {Yasuda}, \& {York}}]{2001AJ....121.2358B}
{Blanton}, M.~R., {Dalcanton}, J., {Eisenstein}, D., {et~al.} 2001, \aj, 121,
  2358

\bibitem[{{Brough} {et~al.}(2002){Brough}, {Collins}, {Burke}, {Mann}, \&
  {Lynam}}]{2002MNRAS.329L..53B}
{Brough}, S., {Collins}, C.~A., {Burke}, D.~J., {Mann}, R.~G., \& {Lynam},
  P.~D. 2002, \mnras, 329, L53

\bibitem[{{Brough} {et~al.}(2008){Brough}, {Couch}, {Collins}, {Jarrett},
  {Burke}, \& {Mann}}]{2008MNRAS.385L.103B}
{Brough}, S., {Couch}, W.~J., {Collins}, C.~A., {et~al.} 2008, \mnras, 385,
  L103

\bibitem[{{Bruzual} \& {Charlot}(2003)}]{2003MNRAS.344.1000B}
{Bruzual}, G., \& {Charlot}, S. 2003, \mnras, 344, 1000

\bibitem[{{Burke} {et~al.}(2012){Burke}, {Collins}, {Stott}, \&
  {Hilton}}]{2012MNRAS.425.2058B}
{Burke}, C., {Collins}, C.~A., {Stott}, J.~P., \& {Hilton}, M. 2012, \mnras,
  425, 2058

\bibitem[{{Burke} {et~al.}(2015){Burke}, {Hilton}, \&
  {Collins}}]{2015arXiv150304321B}
{Burke}, C., {Hilton}, M., \& {Collins}, C. 2015, ArXiv e-prints

\bibitem[{{Busha}(in prep.)}]{bushaprep}
{Busha}, M. in prep., TBD.

\bibitem[{{Chabrier}(2003)}]{2003PASP..115..763C}
{Chabrier}, G. 2003, \pasp, 115, 763

\bibitem[{{Chang} {et~al.}(2014){Chang}, {Busha}, {Wechsler}, {Refregier},
  {Amara}, {Rykoff}, {Becker}, {Bruderer}, {Gamper}, {Leistedt}, {Peiris},
  {Abbott}, {Abdalla}, {Banerji}, {Bernstein}, {Bertin}, {Brooks}, {Carnero
  Rosell}, {Desai}, {da Costa}, {Cunha}, {Eifler}, {Evrard}, {Fausti Neto},
  {Gerdes}, {Gruen}, {James}, {Kuehn}, {Maia}, {Makler}, {Ogando}, {Plazas},
  {Sanchez}, {Schubnell}, {Sevilla-Noarbe}, {Smith}, {Soares-Santos},
  {Suchyta}, {Swanson}, {Tarle}, \& {Zuntz}}]{2014arXiv1411.0032C}
{Chang}, C., {Busha}, M.~T., {Wechsler}, R.~H., {et~al.} 2014, ArXiv e-prints

\bibitem[{{Chiu} {et~al.}(2014){Chiu}, {Mohr}, {Mcdonald}, {Bocquet}, {Ashby},
  {Bayliss}, {Benson}, {Bleem}, {Brodwin}, {Desai}, {Dietrich}, {Forman},
  {Gangkofner}, {Gonzalez}, {Hennig}, {Liu}, {Reichardt}, {Saro}, {Stalder},
  {Stanford}, {Song}, {Schrabback}, {Suhada}, {Strazzullo}, \&
  {Zenteno}}]{2014arXiv1412.7823C}
{Chiu}, I., {Mohr}, J., {Mcdonald}, M., {et~al.} 2014, ArXiv e-prints

\bibitem[{{Collins} {et~al.}(2009){Collins}, {Stott}, {Hilton}, {Kay},
  {Stanford}, {Davidson}, {Hosmer}, {Hoyle}, {Liddle}, {Lloyd-Davies}, {Mann},
  {Mehrtens}, {Miller}, {Nichol}, {Romer}, {Sahl{\'e}n}, {Viana}, \&
  {West}}]{2009Natur.458..603C}
{Collins}, C.~A., {Stott}, J.~P., {Hilton}, M., {et~al.} 2009, \nat, 458, 603

\bibitem[{{Conroy} \& {Gunn}(2010)}]{2010ApJ...712..833C}
{Conroy}, C., \& {Gunn}, J.~E. 2010, \apj, 712, 833

\bibitem[{{Conroy} {et~al.}(2009){Conroy}, {Gunn}, \&
  {White}}]{2009ApJ...699..486C}
{Conroy}, C., {Gunn}, J.~E., \& {White}, M. 2009, \apj, 699, 486

\bibitem[{{Conroy} {et~al.}(2007){Conroy}, {Wechsler}, \&
  {Kravtsov}}]{2007ApJ...668..826C}
{Conroy}, C., {Wechsler}, R.~H., \& {Kravtsov}, A.~V. 2007, \apj, 668, 826

\bibitem[{{Contini} {et~al.}(2014){Contini}, {De Lucia}, {Villalobos}, \&
  {Borgani}}]{2014MNRAS.437.3787C}
{Contini}, E., {De Lucia}, G., {Villalobos}, {\'A}., \& {Borgani}, S. 2014,
  \mnras, 437, 3787

\bibitem[{{Croton} {et~al.}(2006){Croton}, {Springel}, {White}, {De Lucia},
  {Frenk}, {Gao}, {Jenkins}, {Kauffmann}, {Navarro}, \&
  {Yoshida}}]{2006MNRAS.367..864C}
{Croton}, D.~J., {Springel}, V., {White}, S.~D.~M., {et~al.} 2006, \mnras, 367,
  864

\bibitem[{{Cui} {et~al.}(2014){Cui}, {Murante}, {Monaco}, {Borgani}, {Granato},
  {Killedar}, {De Lucia}, {Presotto}, \& {Dolag}}]{2014MNRAS.437..816C}
{Cui}, W., {Murante}, G., {Monaco}, P., {et~al.} 2014, \mnras, 437, 816

\bibitem[{{De Lucia} \& {Blaizot}(2007)}]{2007MNRAS.375....2D}
{De Lucia}, G., \& {Blaizot}, J. 2007, \mnras, 375, 2

\bibitem[{{DeMaio} {et~al.}(2015){DeMaio}, {Gonzalez}, {Zabludoff}, {Zaritsky},
  \& {Bradac}}]{2015arXiv150102251D}
{DeMaio}, T., {Gonzalez}, A., {Zabludoff}, A., {Zaritsky}, D., \& {Bradac}, M.
  2015, ArXiv e-prints

\bibitem[{Diehl {et~al.}(2014)Diehl, Abbott, Annis, Armstrong, Baruah, Bermeo,
  Bernstein, Beynon, Bruderer, Buckley-Geer, {et~al.}}]{diehl2014dark}
Diehl, H., Abbott, T., Annis, J., {et~al.} 2014, in SPIE Astronomical
  Telescopes+ Instrumentation, International Society for Optics and Photonics,
  91490V--91490V

\bibitem[{{Dietrich} {et~al.}(2014){Dietrich}, {Zhang}, {Song}, {Davis},
  {McKay}, {Baruah}, {Becker}, {Benoist}, {Busha}, {da Costa}, {Hao}, {Maia},
  {Miller}, {Ogando}, {Romer}, {Rozo}, {Rykoff}, \&
  {Wechsler}}]{2014MNRAS.443.1713D}
{Dietrich}, J.~P., {Zhang}, Y., {Song}, J., {et~al.} 2014, \mnras, 443, 1713

\bibitem[{{Donahue} {et~al.}(2015){Donahue}, {Connor}, {Fogarty}, {Li}, {Voit},
  {Postman}, {Koekemoer}, {Moustakas}, {Bradley}, \&
  {Ford}}]{2015arXiv150400598D}
{Donahue}, M., {Connor}, T., {Fogarty}, K., {et~al.} 2015, ArXiv e-prints

\bibitem[{{Dubinski}(1998)}]{1998ApJ...502..141D}
{Dubinski}, J. 1998, \apj, 502, 141

\bibitem[{{Eckmiller} {et~al.}(2011){Eckmiller}, {Hudson}, \&
  {Reiprich}}]{2011A&A...535A.105E}
{Eckmiller}, H.~J., {Hudson}, D.~S., \& {Reiprich}, T.~H. 2011, \aap, 535, A105

\bibitem[{{Fabian}(1994)}]{1994ARA&A..32..277F}
{Fabian}, A.~C. 1994, \araa, 32, 277

\bibitem[{{Fabian}(2012)}]{2012ARA&A..50..455F}
---. 2012, \araa, 50, 455

\bibitem[{{Flaugher} {et~al.}(2015){Flaugher}, {Diehl}, {Honscheid}, {Abbott},
  {Alvarez}, {Angstadt}, {Annis}, {Antonik}, {Ballester}, {Beaufore},
  {Bernstein}, {Bernstein}, {Bigelow}, {Bonati}, {Boprie}, {Brooks},
  {Buckley-Geer}, {Campa}, {Cardiel-Sas}, {Castander}, {Castilla}, {Cease},
  {Cela-Ruiz}, {Chappa}, {Chi}, {Cooper}, {da Costa}, {Dede}, {Derylo},
  {DePoy}, {de Vicente}, {Doel}, {Drlica-Wagner}, {Eiting}, {Elliott}, {Emes},
  {Estrada}, {Fausti Neto}, {Finley}, {Flores}, {Frieman}, {Gerdes},
  {Gladders}, {Gregory}, {Gutierrez}, {Hao}, {Holland}, {Holm}, {Huffman},
  {Jackson}, {James}, {Jonas}, {Karcher}, {Karliner}, {Kent}, {Kessler},
  {Kozlovsky}, {Kron}, {Kubik}, {Kuehn}, {Kuhlmann}, {Kuk}, {Lahav}, {Lathrop},
  {Lee}, {Levi}, {Lewis}, {Li}, {Mandrichenko}, {Marshall}, {Martinez},
  {Merritt}, {Miquel}, {Mu{\~n}oz}, {Neilsen}, {Nichol}, {Nord}, {Ogando},
  {Olsen}, {Palaio}, {Patton}, {Peoples}, {Plazas}, {Rauch}, {Reil}, {Rheault},
  {Roe}, {Rogers}, {Roodman}, {Sanchez}, {Scarpine}, {Schindler}, {Schmidt},
  {Schmitt}, {Schubnell}, {Schultz}, {Schurter}, {Scott}, {Serrano}, {Shaw},
  {Smith}, {Soares-Santos}, {Stefanik}, {Stuermer}, {Suchyta}, {Sypniewski},
  {Tarle}, {Thaler}, {Tighe}, {Tran}, {Tucker}, {Walker}, {Wang}, {Watson},
  {Weaverdyck}, {Wester}, {Woods}, {Yanny}, \& {The DES
  Collaboration}}]{2015AJ....150..150F}
{Flaugher}, B., {Diehl}, H.~T., {Honscheid}, K., {et~al.} 2015, \aj, 150, 150

\bibitem[{{Fraser-McKelvie} {et~al.}(2014){Fraser-McKelvie}, {Brown}, \&
  {Pimbblet}}]{2014MNRAS.444L..63F}
{Fraser-McKelvie}, A., {Brown}, M.~J.~I., \& {Pimbblet}, K.~A. 2014, \mnras,
  444, L63

\bibitem[{{Giallongo} {et~al.}(2014){Giallongo}, {Menci}, {Grazian},
  {Gallozzi}, {Castellano}, {Fiore}, {Fontana}, {Pentericci}, {Boutsia},
  {Paris}, {Speziali}, \& {Testa}}]{2014ApJ...781...24G}
{Giallongo}, E., {Menci}, N., {Grazian}, A., {et~al.} 2014, \apj, 781, 24

\bibitem[{{Gonzalez} {et~al.}(2007){Gonzalez}, {Zaritsky}, \&
  {Zabludoff}}]{2007ApJ...666..147G}
{Gonzalez}, A.~H., {Zaritsky}, D., \& {Zabludoff}, A.~I. 2007, \apj, 666, 147

\bibitem[{{Graham} \& {Driver}(2005)}]{2005PASA...22..118G}
{Graham}, A.~W., \& {Driver}, S.~P. 2005, PASA, 22, 118

\bibitem[{{Groenewald} \& {Loubser}(2014)}]{2014MNRAS.444..808G}
{Groenewald}, D.~N., \& {Loubser}, S.~I. 2014, \mnras, 444, 808

\bibitem[{{Guennou} {et~al.}(2012){Guennou}, {Adami}, {Da Rocha}, {Durret},
  {Ulmer}, {Allam}, {Basa}, {Benoist}, {Biviano}, {Clowe}, {Gavazzi},
  {Halliday}, {Ilbert}, {Johnston}, {Just}, {Kron}, {Kubo}, {Le Brun},
  {Marshall}, {Mazure}, {Murphy}, {Pereira}, {Raba{\c c}a}, {Rostagni},
  {Rudnick}, {Russeil}, {Schrabback}, {Slezak}, {Tucker}, \&
  {Zaritsky}}]{2012A&A...537A..64G}
{Guennou}, L., {Adami}, C., {Da Rocha}, C., {et~al.} 2012, \aap, 537, A64

\bibitem[{{Guo} {et~al.}(2013){Guo}, {White}, {Angulo}, {Henriques}, {Lemson},
  {Boylan-Kolchin}, {Thomas}, \& {Short}}]{2013MNRAS.428.1351G}
{Guo}, Q., {White}, S., {Angulo}, R.~E., {et~al.} 2013, \mnras, 428, 1351

\bibitem[{{Guo} {et~al.}(2011){Guo}, {White}, {Boylan-Kolchin}, {De Lucia},
  {Kauffmann}, {Lemson}, {Li}, {Springel}, \& {Weinmann}}]{2011MNRAS.413..101G}
{Guo}, Q., {White}, S., {Boylan-Kolchin}, M., {et~al.} 2011, \mnras, 413, 101

\bibitem[{{Harrison} {et~al.}(2012){Harrison}, {Miller}, {Richards},
  {Lloyd-Davies}, {Hoyle}, {Romer}, {Mehrtens}, {Hilton}, {Stott}, {Capozzi},
  {Collins}, {Deadman}, {Liddle}, {Sahl{\'e}n}, {Stanford}, \&
  {Viana}}]{2012ApJ...752...12H}
{Harrison}, C.~D., {Miller}, C.~J., {Richards}, J.~W., {et~al.} 2012, \apj,
  752, 12

\bibitem[{{Hausman} \& {Ostriker}(1978)}]{1978ApJ...224..320H}
{Hausman}, M.~A., \& {Ostriker}, J.~P. 1978, \apj, 224, 320

\bibitem[{{H{\"a}ussler} {et~al.}(2007){H{\"a}ussler}, {McIntosh}, {Barden},
  {Bell}, {Rix}, {Borch}, {Beckwith}, {Caldwell}, {Heymans}, {Jahnke}, {Jogee},
  {Koposov}, {Meisenheimer}, {S{\'a}nchez}, {Somerville}, {Wisotzki}, \&
  {Wolf}}]{2007ApJS..172..615H}
{H{\"a}ussler}, B., {McIntosh}, D.~H., {Barden}, M., {et~al.} 2007, \apjs, 172,
  615

\bibitem[{{He} {et~al.}(2013){He}, {Xia}, {Hao}, {Jing}, {Mao}, \&
  {Li}}]{2013ApJ...773...37H}
{He}, Y.~Q., {Xia}, X.~Y., {Hao}, C.~N., {et~al.} 2013, \apj, 773, 37

\bibitem[{{Henry} {et~al.}(2009){Henry}, {Evrard}, {Hoekstra}, {Babul}, \&
  {Mahdavi}}]{2009ApJ...691.1307H}
{Henry}, J.~P., {Evrard}, A.~E., {Hoekstra}, H., {Babul}, A., \& {Mahdavi}, A.
  2009, \apj, 691, 1307

\bibitem[{{High} {et~al.}(2009){High}, {Stubbs}, {Rest}, {Stalder}, \&
  {Challis}}]{2009AJ....138..110H}
{High}, F.~W., {Stubbs}, C.~W., {Rest}, A., {Stalder}, B., \& {Challis}, P.
  2009, \aj, 138, 110

\bibitem[{{Hilton} {et~al.}(2012){Hilton}, {Romer}, {Kay}, {Mehrtens},
  {Lloyd-Davies}, {Thomas}, {Short}, {Mayers}, {Rooney}, {Stott}, {Collins},
  {Harrison}, {Hoyle}, {Liddle}, {Mann}, {Miller}, {Sahl{\'e}n}, {Viana},
  {Davidson}, {Hosmer}, {Nichol}, {Sabirli}, {Stanford}, \&
  {West}}]{2012MNRAS.424.2086H}
{Hilton}, M., {Romer}, A.~K., {Kay}, S.~T., {et~al.} 2012, \mnras, 424, 2086

\bibitem[{{Hoekstra} {et~al.}(2011){Hoekstra}, {Donahue}, {Conselice},
  {McNamara}, \& {Voit}}]{2011ApJ...726...48H}
{Hoekstra}, H., {Donahue}, M., {Conselice}, C.~J., {McNamara}, B.~R., \&
  {Voit}, G.~M. 2011, \apj, 726, 48

\bibitem[{{Hoekstra} {et~al.}(2012){Hoekstra}, {Mahdavi}, {Babul}, \&
  {Bildfell}}]{2012MNRAS.427.1298H}
{Hoekstra}, H., {Mahdavi}, A., {Babul}, A., \& {Bildfell}, C. 2012, \mnras,
  427, 1298

\bibitem[{{Hogg} {et~al.}(2010){Hogg}, {Bovy}, \& {Lang}}]{2010arXiv1008.4686H}
{Hogg}, D.~W., {Bovy}, J., \& {Lang}, D. 2010, ArXiv e-prints

\bibitem[{{Hu} \& {Kravtsov}(2003)}]{2003ApJ...584..702H}
{Hu}, W., \& {Kravtsov}, A.~V. 2003, \apj, 584, 702

\bibitem[{{Inagaki} {et~al.}(2015){Inagaki}, {Lin}, {Huang}, {Hsieh}, \&
  {Sugiyama}}]{2015MNRAS.446.1107I}
{Inagaki}, T., {Lin}, Y.-T., {Huang}, H.-J., {Hsieh}, B.-C., \& {Sugiyama}, N.
  2015, \mnras, 446, 1107

\bibitem[{{Kelly} {et~al.}(2014){Kelly}, {von der Linden}, {Applegate},
  {Allen}, {Allen}, {Burchat}, {Burke}, {Ebeling}, {Capak}, {Czoske},
  {Donovan}, {Mantz}, \& {Morris}}]{2014MNRAS.439...28K}
{Kelly}, P.~L., {von der Linden}, A., {Applegate}, D.~E., {et~al.} 2014,
  \mnras, 439, 28

\bibitem[{{Kettula} {et~al.}(2013){Kettula}, {Finoguenov}, {Massey}, {Rhodes},
  {Hoekstra}, {Taylor}, {Spinelli}, {Tanaka}, {Ilbert}, {Capak}, {McCracken},
  \& {Koekemoer}}]{2013ApJ...778...74K}
{Kettula}, K., {Finoguenov}, A., {Massey}, R., {et~al.} 2013, \apj, 778, 74

\bibitem[{{Kravtsov} {et~al.}(2014){Kravtsov}, {Vikhlinin}, \&
  {Meshscheryakov}}]{2014arXiv1401.7329K}
{Kravtsov}, A., {Vikhlinin}, A., \& {Meshscheryakov}, A. 2014, ArXiv e-prints

\bibitem[{{Kravtsov} {et~al.}(2006){Kravtsov}, {Vikhlinin}, \&
  {Nagai}}]{2006ApJ...650..128K}
{Kravtsov}, A.~V., {Vikhlinin}, A., \& {Nagai}, D. 2006, \apj, 650, 128

\bibitem[{{Krick} \& {Bernstein}(2007)}]{2007AJ....134..466K}
{Krick}, J.~E., \& {Bernstein}, R.~A. 2007, \aj, 134, 466

\bibitem[{{Krick} {et~al.}(2006){Krick}, {Bernstein}, \&
  {Pimbblet}}]{2006AJ....131..168K}
{Krick}, J.~E., {Bernstein}, R.~A., \& {Pimbblet}, K.~A. 2006, \aj, 131, 168

\bibitem[{{Kron}(1980)}]{1980ApJS...43..305K}
{Kron}, R.~G. 1980, \apjs, 43, 305

\bibitem[{{Kroupa}(2001)}]{2001MNRAS.322..231K}
{Kroupa}, P. 2001, \mnras, 322, 231

\bibitem[{{Laporte} {et~al.}(2013){Laporte}, {White}, {Naab}, \&
  {Gao}}]{2013MNRAS.435..901L}
{Laporte}, C.~F.~P., {White}, S.~D.~M., {Naab}, T., \& {Gao}, L. 2013, \mnras,
  435, 901

\bibitem[{{Lauer} {et~al.}(2007){Lauer}, {Faber}, {Richstone}, {Gebhardt},
  {Tremaine}, {Postman}, {Dressler}, {Aller}, {Filippenko}, {Green}, {Ho},
  {Kormendy}, {Magorrian}, \& {Pinkney}}]{2007ApJ...662..808L}
{Lauer}, T.~R., {Faber}, S.~M., {Richstone}, D., {et~al.} 2007, \apj, 662, 808

\bibitem[{{Li} \& {Bryan}(2014{\natexlab{a}})}]{2014ApJ...789...54L}
{Li}, Y., \& {Bryan}, G.~L. 2014{\natexlab{a}}, \apj, 789, 54

\bibitem[{{Li} \& {Bryan}(2014{\natexlab{b}})}]{2014ApJ...789..153L}
---. 2014{\natexlab{b}}, \apj, 789, 153

\bibitem[{{Lidman} {et~al.}(2012){Lidman}, {Suherli}, {Muzzin}, {Wilson},
  {Demarco}, {Brough}, {Rettura}, {Cox}, {DeGroot}, {Yee}, {Gilbank},
  {Hoekstra}, {Balogh}, {Ellingson}, {Hicks}, {Nantais}, {Noble}, {Lacy},
  {Surace}, \& {Webb}}]{2012MNRAS.427..550L}
{Lidman}, C., {Suherli}, J., {Muzzin}, A., {et~al.} 2012, \mnras, 427, 550

\bibitem[{{Lin} {et~al.}(2013{\natexlab{a}}){Lin}, {Soares-Santos}, {Diehl}, \&
  {Dark Energy Survey Collaboration}}]{2013AAS...22135226L}
{Lin}, H., {Soares-Santos}, M., {Diehl}, H., \& {Dark Energy Survey
  Collaboration}. 2013{\natexlab{a}}, in American Astronomical Society Meeting
  Abstracts, Vol. 221, American Astronomical Society Meeting Abstracts \#221,
  352.26

\bibitem[{{Lin} {et~al.}(2013{\natexlab{b}}){Lin}, {Brodwin}, {Gonzalez},
  {Bode}, {Eisenhardt}, {Stanford}, \& {Vikhlinin}}]{2013ApJ...771...61L}
{Lin}, Y.-T., {Brodwin}, M., {Gonzalez}, A.~H., {et~al.} 2013{\natexlab{b}},
  \apj, 771, 61

\bibitem[{{Lin} \& {Mohr}(2004)}]{2004ApJ...617..879L}
{Lin}, Y.-T., \& {Mohr}, J.~J. 2004, \apj, 617, 879

\bibitem[{{Liu} {et~al.}(2012){Liu}, {Mao}, \& {Meng}}]{2012MNRAS.423..422L}
{Liu}, F.~S., {Mao}, S., \& {Meng}, X.~M. 2012, \mnras, 423, 422

\bibitem[{{Liu} {et~al.}(2013){Liu}, {Guo}, {Koo}, {Trump}, {Barro}, {Yesuf},
  {Faber}, {Giavalisco}, {Cassata}, {Koekemoer}, {Pentericci}, {Castellano},
  {Cheung}, {Mao}, {Xia}, {Grogin}, {Hathi}, {Huang}, {Kocevski}, {McGrath}, \&
  {Wuyts}}]{2013ApJ...769..147L}
{Liu}, F.~S., {Guo}, Y., {Koo}, D.~C., {et~al.} 2013, \apj, 769, 147

\bibitem[{{Lloyd-Davies} {et~al.}(2011){Lloyd-Davies}, {Romer}, {Mehrtens},
  {Hosmer}, {Davidson}, {Sabirli}, {Mann}, {Hilton}, {Liddle}, {Viana},
  {Campbell}, {Collins}, {Dubois}, {Freeman}, {Harrison}, {Hoyle}, {Kay},
  {Kuwertz}, {Miller}, {Nichol}, {Sahl{\'e}n}, {Stanford}, \&
  {Stott}}]{2011MNRAS.418...14L}
{Lloyd-Davies}, E.~J., {Romer}, A.~K., {Mehrtens}, N., {et~al.} 2011, \mnras,
  418, 14

\bibitem[{{Lu} {et~al.}(2014){Lu}, {Mo}, {Lu}, {Katz}, {Weinberg}, {van den
  Bosch}, \& {Yang}}]{2014MNRAS.439.1294L}
{Lu}, Z., {Mo}, H.~J., {Lu}, Y., {et~al.} 2014, \mnras, 439, 1294

\bibitem[{{Mahdavi} {et~al.}(2013){Mahdavi}, {Hoekstra}, {Babul}, {Bildfell},
  {Jeltema}, \& {Henry}}]{2013ApJ...767..116M}
{Mahdavi}, A., {Hoekstra}, H., {Babul}, A., {et~al.} 2013, \apj, 767, 116

\bibitem[{{Mancone} \& {Gonzalez}(2012)}]{2012PASP..124..606M}
{Mancone}, C.~L., \& {Gonzalez}, A.~H. 2012, \pasp, 124, 606

\bibitem[{{Mandelbaum} {et~al.}(2005){Mandelbaum}, {Hirata}, {Seljak}, {Guzik},
  {Padmanabhan}, {Blake}, {Blanton}, {Lupton}, \&
  {Brinkmann}}]{2005MNRAS.361.1287M}
{Mandelbaum}, R., {Hirata}, C.~M., {Seljak}, U., {et~al.} 2005, \mnras, 361,
  1287

\bibitem[{{Mantz} {et~al.}(2010){Mantz}, {Allen}, {Ebeling}, {Rapetti}, \&
  {Drlica-Wagner}}]{2010MNRAS.406.1773M}
{Mantz}, A., {Allen}, S.~W., {Ebeling}, H., {Rapetti}, D., \& {Drlica-Wagner},
  A. 2010, \mnras, 406, 1773

\bibitem[{{Maraston}(2005)}]{2005MNRAS.362..799M}
{Maraston}, C. 2005, \mnras, 362, 799

\bibitem[{{Martizzi} {et~al.}(2014){Martizzi}, {Jimmy}, {Teyssier}, \&
  {Moore}}]{2014MNRAS.443.1500M}
{Martizzi}, D., {Jimmy}, {Teyssier}, R., \& {Moore}, B. 2014, \mnras, 443, 1500

\bibitem[{{Martizzi} {et~al.}(2012){Martizzi}, {Teyssier}, \&
  {Moore}}]{2012MNRAS.420.2859M}
{Martizzi}, D., {Teyssier}, R., \& {Moore}, B. 2012, \mnras, 420, 2859

\bibitem[{{Matthews} {et~al.}(1964){Matthews}, {Morgan}, \&
  {Schmidt}}]{1964ApJ...140...35M}
{Matthews}, T.~A., {Morgan}, W.~W., \& {Schmidt}, M. 1964, \apj, 140, 35

\bibitem[{{Maughan} {et~al.}(2012){Maughan}, {Giles}, {Randall}, {Jones}, \&
  {Forman}}]{2012MNRAS.421.1583M}
{Maughan}, B.~J., {Giles}, P.~A., {Randall}, S.~W., {Jones}, C., \& {Forman},
  W.~R. 2012, \mnras, 421, 1583

\bibitem[{{McDonald} {et~al.}(2012){McDonald}, {Bayliss}, {Benson}, {Foley},
  {Ruel}, {Sullivan}, {Veilleux}, {Aird}, {Ashby}, {Bautz}, {Bazin}, {Bleem},
  {Brodwin}, {Carlstrom}, {Chang}, {Cho}, {Clocchiatti}, {Crawford}, {Crites},
  {de Haan}, {Desai}, {Dobbs}, {Dudley}, {Egami}, {Forman}, {Garmire},
  {George}, {Gladders}, {Gonzalez}, {Halverson}, {Harrington}, {High},
  {Holder}, {Holzapfel}, {Hoover}, {Hrubes}, {Jones}, {Joy}, {Keisler}, {Knox},
  {Lee}, {Leitch}, {Liu}, {Lueker}, {Luong-van}, {Mantz}, {Marrone}, {McMahon},
  {Mehl}, {Meyer}, {Miller}, {Mocanu}, {Mohr}, {Montroy}, {Murray}, {Natoli},
  {Padin}, {Plagge}, {Pryke}, {Rawle}, {Reichardt}, {Rest}, {Rex}, {Ruhl},
  {Saliwanchik}, {Saro}, {Sayre}, {Schaffer}, {Shaw}, {Shirokoff}, {Simcoe},
  {Song}, {Spieler}, {Stalder}, {Staniszewski}, {Stark}, {Story}, {Stubbs},
  {{\v S}uhada}, {van Engelen}, {Vanderlinde}, {Vieira}, {Vikhlinin},
  {Williamson}, {Zahn}, \& {Zenteno}}]{2012Natur.488..349M}
{McDonald}, M., {Bayliss}, M., {Benson}, B.~A., {et~al.} 2012, \nat, 488, 349

\bibitem[{{Meece} {et~al.}(2015){Meece}, {O'Shea}, \&
  {Voit}}]{2015arXiv150302645M}
{Meece}, G., {O'Shea}, B., \& {Voit}, M. 2015, ArXiv e-prints

\bibitem[{{Meert} {et~al.}(2015){Meert}, {Vikram}, \&
  {Bernardi}}]{2015MNRAS.446.3943M}
{Meert}, A., {Vikram}, V., \& {Bernardi}, M. 2015, \mnras, 446, 3943

\bibitem[{{Mehrtens} {et~al.}(2012){Mehrtens}, {Romer}, {Hilton},
  {Lloyd-Davies}, {Miller}, {Stanford}, {Hosmer}, {Hoyle}, {Collins}, {Liddle},
  {Viana}, {Nichol}, {Stott}, {Dubois}, {Kay}, {Sahl{\'e}n}, {Young}, {Short},
  {Christodoulou}, {Watson}, {Davidson}, {Harrison}, {Baruah}, {Smith},
  {Burke}, {Mayers}, {Deadman}, {Rooney}, {Edmondson}, {West}, {Campbell},
  {Edge}, {Mann}, {Sabirli}, {Wake}, {Benoist}, {da Costa}, {Maia}, \&
  {Ogando}}]{2012MNRAS.423.1024M}
{Mehrtens}, N., {Romer}, A.~K., {Hilton}, M., {et~al.} 2012, \mnras, 423, 1024

\bibitem[{{Miller} {et~al.}(in prep.){Miller}, {Rooney}, \& {The DES
  collaboration}}]{2014Miller}
{Miller}, C.~J., {Rooney}, P., \& {The DES collaboration}. in prep., TBD.

\bibitem[{{Mitchell} {et~al.}(2013){Mitchell}, {Lacey}, {Baugh}, \&
  {Cole}}]{2013MNRAS.435...87M}
{Mitchell}, P.~D., {Lacey}, C.~G., {Baugh}, C.~M., \& {Cole}, S. 2013, \mnras,
  435, 87

\bibitem[{{Mohr} {et~al.}(2012){Mohr}, {Armstrong}, {Bertin}, {Daues}, {Desai},
  {Gower}, {Gruendl}, {Hanlon}, {Kuropatkin}, {Lin}, {Marriner}, {Petravic},
  {Sevilla}, {Swanson}, {Tomashek}, {Tucker}, \& {Yanny}}]{2012SPIE.8451E..0DM}
{Mohr}, J.~J., {Armstrong}, R., {Bertin}, E., {et~al.} 2012, in Society of
  Photo-Optical Instrumentation Engineers (SPIE) Conference Series, Vol. 8451,
  Society of Photo-Optical Instrumentation Engineers (SPIE) Conference Series

\bibitem[{{Monaco} {et~al.}(2006){Monaco}, {Murante}, {Borgani}, \&
  {Fontanot}}]{2006ApJ...652L..89M}
{Monaco}, P., {Murante}, G., {Borgani}, S., \& {Fontanot}, F. 2006, \apjl, 652,
  L89

\bibitem[{{Montes} \& {Trujillo}(2014)}]{2014ApJ...794..137M}
{Montes}, M., \& {Trujillo}, I. 2014, \apj, 794, 137

\bibitem[{{Moster} {et~al.}(2013){Moster}, {Naab}, \&
  {White}}]{2013MNRAS.428.3121M}
{Moster}, B.~P., {Naab}, T., \& {White}, S.~D.~M. 2013, \mnras, 428, 3121

\bibitem[{{Moster} {et~al.}(2010){Moster}, {Somerville}, {Maulbetsch}, {van den
  Bosch}, {Macci{\`o}}, {Naab}, \& {Oser}}]{2010ApJ...710..903M}
{Moster}, B.~P., {Somerville}, R.~S., {Maulbetsch}, C., {et~al.} 2010, \apj,
  710, 903

\bibitem[{{Nagai} {et~al.}(2007){Nagai}, {Vikhlinin}, \&
  {Kravtsov}}]{2007ApJ...655...98N}
{Nagai}, D., {Vikhlinin}, A., \& {Kravtsov}, A.~V. 2007, \apj, 655, 98

\bibitem[{{Oliva-Altamirano} {et~al.}(2014){Oliva-Altamirano}, {Brough},
  {Lidman}, {Couch}, {Hopkins}, {Colless}, {Taylor}, {Robotham},
  {Gunawardhana}, {Ponman}, {Baldry}, {Bauer}, {Bland-Hawthorn}, {Cluver},
  {Cameron}, {Conselice}, {Driver}, {Edge}, {Graham}, {van Kampen},
  {Lara-L{\'o}pez}, {Liske}, {L{\'o}pez-S{\'a}nchez}, {Loveday}, {Mahajan},
  {Peacock}, {Phillipps}, {Pimbblet}, \& {Sharp}}]{2014MNRAS.440..762O}
{Oliva-Altamirano}, P., {Brough}, S., {Lidman}, C., {et~al.} 2014, \mnras, 440,
  762

\bibitem[{{Ostriker} \& {Tremaine}(1975)}]{1975ApJ...202L.113O}
{Ostriker}, J.~P., \& {Tremaine}, S.~D. 1975, \apjl, 202, L113

\bibitem[{{Papovich} {et~al.}(2011){Papovich}, {Finkelstein}, {Ferguson},
  {Lotz}, \& {Giavalisco}}]{2011MNRAS.412.1123P}
{Papovich}, C., {Finkelstein}, S.~L., {Ferguson}, H.~C., {Lotz}, J.~M., \&
  {Giavalisco}, M. 2011, \mnras, 412, 1123

\bibitem[{{Patel} {et~al.}(2013){Patel}, {van Dokkum}, {Franx}, {Quadri},
  {Muzzin}, {Marchesini}, {Williams}, {Holden}, \&
  {Stefanon}}]{2013ApJ...766...15P}
{Patel}, S.~G., {van Dokkum}, P.~G., {Franx}, M., {et~al.} 2013, \apj, 766, 15

\bibitem[{{Peng} {et~al.}(2002){Peng}, {Ho}, {Impey}, \&
  {Rix}}]{2002AJ....124..266P}
{Peng}, C.~Y., {Ho}, L.~C., {Impey}, C.~D., \& {Rix}, H.-W. 2002, \aj, 124, 266

\bibitem[{{Peng} {et~al.}(2010){Peng}, {Ho}, {Impey}, \&
  {Rix}}]{2010AJ....139.2097P}
---. 2010, \aj, 139, 2097

\bibitem[{{Petrosian}(1976)}]{1976ApJ...209L...1P}
{Petrosian}, V. 1976, \apjl, 209, L1

\bibitem[{{Pike} {et~al.}(2014){Pike}, {Kay}, {Newton}, {Thomas}, \&
  {Jenkins}}]{2014MNRAS.445.1774P}
{Pike}, S.~R., {Kay}, S.~T., {Newton}, R.~D.~A., {Thomas}, P.~A., \& {Jenkins},
  A. 2014, \mnras, 445, 1774

\bibitem[{{Pratt} {et~al.}(2009){Pratt}, {Croston}, {Arnaud}, \&
  {B{\"o}hringer}}]{2009A&A...498..361P}
{Pratt}, G.~W., {Croston}, J.~H., {Arnaud}, M., \& {B{\"o}hringer}, H. 2009,
  \aap, 498, 361

\bibitem[{{Presotto} {et~al.}(2014){Presotto}, {Girardi}, {Nonino}, {Mercurio},
  {Grillo}, {Rosati}, {Biviano}, {Annunziatella}, {Balestra}, {Cui},
  {Sartoris}, {Lemze}, {Ascaso}, {Moustakas}, {Ford}, {Fritz}, {Czoske},
  {Ettori}, {Kuchner}, {Lombardi}, {Maier}, {Medezinski}, {Molino},
  {Scodeggio}, {Strazzullo}, {Tozzi}, {Ziegler}, {Bartelmann}, {Benitez},
  {Bradley}, {Brescia}, {Broadhurst}, {Coe}, {Donahue}, {Gobat}, {Graves},
  {Kelson}, {Koekemoer}, {Melchior}, {Meneghetti}, {Merten}, {Moustakas},
  {Munari}, {Postman}, {Reg{\H o}s}, {Seitz}, {Umetsu}, {Zheng}, \&
  {Zitrin}}]{2014A&A...565A.126P}
{Presotto}, V., {Girardi}, M., {Nonino}, M., {et~al.} 2014, \aap, 565, A126

\bibitem[{{Puchwein} {et~al.}(2010){Puchwein}, {Springel}, {Sijacki}, \&
  {Dolag}}]{2010MNRAS.406..936P}
{Puchwein}, E., {Springel}, V., {Sijacki}, D., \& {Dolag}, K. 2010, \mnras,
  406, 936

\bibitem[{{Ragone-Figueroa} {et~al.}(2013){Ragone-Figueroa}, {Granato},
  {Murante}, {Borgani}, \& {Cui}}]{2013MNRAS.436.1750R}
{Ragone-Figueroa}, C., {Granato}, G.~L., {Murante}, G., {Borgani}, S., \&
  {Cui}, W. 2013, \mnras, 436, 1750

\bibitem[{{Rasia} {et~al.}(2013){Rasia}, {Borgani}, {Ettori}, {Mazzotta}, \&
  {Meneghetti}}]{2013ApJ...776...39R}
{Rasia}, E., {Borgani}, S., {Ettori}, S., {Mazzotta}, P., \& {Meneghetti}, M.
  2013, \apj, 776, 39

\bibitem[{{Rasia} {et~al.}(2012){Rasia}, {Meneghetti}, {Martino}, {Borgani},
  {Bonafede}, {Dolag}, {Ettori}, {Fabjan}, {Giocoli}, {Mazzotta}, {Merten},
  {Radovich}, \& {Tornatore}}]{2012NJPh...14e5018R}
{Rasia}, E., {Meneghetti}, M., {Martino}, R., {et~al.} 2012, New Journal of
  Physics, 14, 055018

\bibitem[{{Richstone} \& {Malumuth}(1983)}]{1983ApJ...268...30R}
{Richstone}, D.~O., \& {Malumuth}, E.~M. 1983, \apj, 268, 30

\bibitem[{{Rudick} {et~al.}(2011){Rudick}, {Mihos}, \&
  {McBride}}]{2011ApJ...732...48R}
{Rudick}, C.~S., {Mihos}, J.~C., \& {McBride}, C.~K. 2011, \apj, 732, 48

\bibitem[{{Ruszkowski} \& {Springel}(2009)}]{2009ApJ...696.1094R}
{Ruszkowski}, M., \& {Springel}, V. 2009, \apj, 696, 1094

\bibitem[{{Rykoff} {et~al.}(in prep.){Rykoff}, {Bechtol}, \& {DES
  Collaboration}}]{Rykoff}
{Rykoff}, E., {Bechtol}, R., \& {DES Collaboration}. in prep., TBD

\bibitem[{{Rykoff} {et~al.}(2014){Rykoff}, {Rozo}, {Busha}, {Cunha},
  {Finoguenov}, {Evrard}, {Hao}, {Koester}, {Leauthaud}, {Nord}, {Pierre},
  {Reddick}, {Sadibekova}, {Sheldon}, \& {Wechsler}}]{2014ApJ...785..104R}
{Rykoff}, E.~S., {Rozo}, E., {Busha}, M.~T., {et~al.} 2014, \apj, 785, 104

\bibitem[{{S{\'a}nchez} {et~al.}(2014){S{\'a}nchez}, {Carrasco Kind}, {Lin},
  {Miquel}, {Abdalla}, {Amara}, {Banerji}, {Bonnett}, {Brunner}, {Capozzi},
  {Carnero}, {Castander}, {da Costa}, {Cunha}, {Fausti}, {Gerdes}, {Greisel},
  {Gschwend}, {Hartley}, {Jouvel}, {Lahav}, {Lima}, {Maia}, {Mart{\'{\i}}},
  {Ogando}, {Ostrovski}, {Pellegrini}, {Rau}, {Sadeh}, {Seitz},
  {Sevilla-Noarbe}, {Sypniewski}, {de Vicente}, {Abbot}, {Allam}, {Atlee},
  {Bernstein}, {Bernstein}, {Buckley-Geer}, {Burke}, {Childress}, {Davis},
  {DePoy}, {Dey}, {Desai}, {Diehl}, {Doel}, {Estrada}, {Evrard},
  {Fern{\'a}ndez}, {Finley}, {Flaugher}, {Frieman}, {Gaztanaga}, {Glazebrook},
  {Honscheid}, {Kim}, {Kuehn}, {Kuropatkin}, {Lidman}, {Makler}, {Marshall},
  {Nichol}, {Roodman}, {S{\'a}nchez}, {Santiago}, {Sako}, {Scalzo}, {Smith},
  {Swanson}, {Tarle}, {Thomas}, {Tucker}, {Uddin}, {Vald{\'e}s}, {Walker},
  {Yuan}, \& {Zuntz}}]{2014MNRAS.445.1482S}
{S{\'a}nchez}, C., {Carrasco Kind}, M., {Lin}, H., {et~al.} 2014, \mnras, 445,
  1482

\bibitem[{{S{\'a}nchez}(2010)}]{2010JPhCS.259a2080S}
{S{\'a}nchez}, E. 2010, Journal of Physics Conference Series, 259, 012080

\bibitem[{{Shankar} {et~al.}(2015){Shankar}, {Buchan}, {Rettura}, {Bouillot},
  {Moreno}, {Licitra}, {Bernardi}, {Huertas-Company}, {Mei}, {Ascaso}, {Sheth},
  {Delaye}, \& {Raichoor}}]{2015arXiv150102800S}
{Shankar}, F., {Buchan}, S., {Rettura}, A., {et~al.} 2015, ArXiv e-prints

\bibitem[{{Springel} {et~al.}(2005){Springel}, {White}, {Jenkins}, {Frenk},
  {Yoshida}, {Gao}, {Navarro}, {Thacker}, {Croton}, {Helly}, {Peacock}, {Cole},
  {Thomas}, {Couchman}, {Evrard}, {Colberg}, \& {Pearce}}]{2005Natur.435..629S}
{Springel}, V., {White}, S.~D.~M., {Jenkins}, A., {et~al.} 2005, \nat, 435, 629

\bibitem[{{Stott} {et~al.}(2011){Stott}, {Collins}, {Burke}, {Hamilton-Morris},
  \& {Smith}}]{2011MNRAS.414..445S}
{Stott}, J.~P., {Collins}, C.~A., {Burke}, C., {Hamilton-Morris}, V., \&
  {Smith}, G.~P. 2011, \mnras, 414, 445

\bibitem[{{Stott} {et~al.}(2010){Stott}, {Collins}, {Sahl{\'e}n}, {Hilton},
  {Lloyd-Davies}, {Capozzi}, {Hosmer}, {Liddle}, {Mehrtens}, {Miller}, {Romer},
  {Stanford}, {Viana}, {Davidson}, {Hoyle}, {Kay}, \&
  {Nichol}}]{2010ApJ...718...23S}
{Stott}, J.~P., {Collins}, C.~A., {Sahl{\'e}n}, M., {et~al.} 2010, \apj, 718,
  23

\bibitem[{{Stott} {et~al.}(2012){Stott}, {Hickox}, {Edge}, {Collins}, {Hilton},
  {Harrison}, {Romer}, {Rooney}, {Kay}, {Miller}, {Sahl{\'e}n}, {Lloyd-Davies},
  {Mehrtens}, {Hoyle}, {Liddle}, {Viana}, {McCarthy}, {Schaye}, \&
  {Booth}}]{2012MNRAS.422.2213S}
{Stott}, J.~P., {Hickox}, R.~C., {Edge}, A.~C., {et~al.} 2012, \mnras, 422,
  2213

\bibitem[{{Suchyta} {et~al.}(2015){Suchyta}, {Huff}, {Aleksi{\'c}}, {Melchior},
  {Jouvel}, {MacCrann}, {Crocce}, {Gaztanaga}, {Honscheid}, {Leistedt},
  {Peiris}, {Ross}, {Rykoff}, {Sheldon}, {Abbott}, {Abdalla}, {Allam},
  {Banerji}, {Benoit-L{\'e}vy}, {Bertin}, {Brooks}, {Burke}, {Carnero Rosell},
  {Carrasco Kind}, {Carretero}, {Cunha}, {D'Andrea}, {da Costa}, {DePoy},
  {Desai}, {Diehl}, {Dietrich}, {Doel}, {Eifler}, {Estrada}, {Evrard},
  {Flaugher}, {Fosalba}, {Frieman}, {Gerdes}, {Gruen}, {Gruendl}, {James},
  {Jarvis}, {Kuehn}, {Kuropatkin}, {Lahav}, {Lima}, {Maia}, {March},
  {Marshall}, {Miller}, {Miquel}, {Neilsen}, {Nichol}, {Nord}, {Ogando},
  {Percival}, {Reil}, {Roodman}, {Sako}, {Sanchez}, {Scarpine},
  {Sevilla-Noarbe}, {Smith}, {Soares-Santos}, {Sobreira}, {Swanson}, {Tarle},
  {Thaler}, {Thomas}, {Vikram}, {Walker}, {Wechsler}, \&
  {Zhang}}]{2015arXiv150708336S}
{Suchyta}, E., {Huff}, E.~M., {Aleksi{\'c}}, J., {et~al.} 2015, ArXiv e-prints

\bibitem[{{Sun} {et~al.}(2009){Sun}, {Voit}, {Donahue}, {Jones}, {Forman}, \&
  {Vikhlinin}}]{2009ApJ...693.1142S}
{Sun}, M., {Voit}, G.~M., {Donahue}, M., {et~al.} 2009, \apj, 693, 1142

\bibitem[{{Toledo} {et~al.}(2011){Toledo}, {Melnick}, {Selman}, {Quintana},
  {Giraud}, \& {Zelaya}}]{2011MNRAS.414..602T}
{Toledo}, I., {Melnick}, J., {Selman}, F., {et~al.} 2011, \mnras, 414, 602

\bibitem[{{Tonini} {et~al.}(2012){Tonini}, {Bernyk}, {Croton}, {Maraston}, \&
  {Thomas}}]{2012ApJ...759...43T}
{Tonini}, C., {Bernyk}, M., {Croton}, D., {Maraston}, C., \& {Thomas}, D. 2012,
  \apj, 759, 43

\bibitem[{{Tremaine} \& {Richstone}(1977)}]{1977ApJ...212..311T}
{Tremaine}, S.~D., \& {Richstone}, D.~O. 1977, \apj, 212, 311

\bibitem[{{van der Burg} {et~al.}(2014){van der Burg}, {Muzzin}, {Hoekstra},
  {Wilson}, {Lidman}, \& {Yee}}]{2014A&A...561A..79V}
{van der Burg}, R.~F.~J., {Muzzin}, A., {Hoekstra}, H., {et~al.} 2014, \aap,
  561, A79

\bibitem[{{van Dokkum} {et~al.}(2010){van Dokkum}, {Whitaker}, {Brammer},
  {Franx}, {Kriek}, {Labb{\'e}}, {Marchesini}, {Quadri}, {Bezanson},
  {Illingworth}, {Muzzin}, {Rudnick}, {Tal}, \& {Wake}}]{2010ApJ...709.1018V}
{van Dokkum}, P.~G., {Whitaker}, K.~E., {Brammer}, G., {et~al.} 2010, \apj,
  709, 1018

\bibitem[{{Viana} {et~al.}(2013){Viana}, {Mehrtens}, {Harrison}, {Romer},
  {Collins}, {Hilton}, {Hoyle}, {Kay}, {Liddle}, {Mayers}, {Miller}, {Rooney},
  {Sahl{\'e}n}, \& {Stott}}]{2013AN....334..462V}
{Viana}, P.~T.~P., {Mehrtens}, N., {Harrison}, C.~D., {et~al.} 2013,
  Astronomische Nachrichten, 334, 462

\bibitem[{{Vikhlinin} {et~al.}(2009){Vikhlinin}, {Burenin}, {Ebeling},
  {Forman}, {Hornstrup}, {Jones}, {Kravtsov}, {Murray}, {Nagai}, {Quintana}, \&
  {Voevodkin}}]{2009ApJ...692.1033V}
{Vikhlinin}, A., {Burenin}, R.~A., {Ebeling}, H., {et~al.} 2009, \apj, 692,
  1033

\bibitem[{{Voit} {et~al.}(2015){Voit}, {Donahue}, {Bryan}, \&
  {McDonald}}]{2015Natur.519..203V}
{Voit}, G.~M., {Donahue}, M., {Bryan}, G.~L., \& {McDonald}, M. 2015, \nat,
  519, 203

\bibitem[{{von der Linden} {et~al.}(2007){von der Linden}, {Best}, {Kauffmann},
  \& {White}}]{2007MNRAS.379..867V}
{von der Linden}, A., {Best}, P.~N., {Kauffmann}, G., \& {White}, S.~D.~M.
  2007, \mnras, 379, 867

\bibitem[{{Wechsler}(in prep.)}]{weschlerprep}
{Wechsler}, R. in prep., TBD.

\bibitem[{{Whiley} {et~al.}(2008){Whiley}, {Arag{\'o}n-Salamanca}, {De Lucia},
  {von der Linden}, {Bamford}, {Best}, {Bremer}, {Jablonka}, {Johnson},
  {Milvang-Jensen}, {Noll}, {Poggianti}, {Rudnick}, {Saglia}, {White}, \&
  {Zaritsky}}]{2008MNRAS.387.1253W}
{Whiley}, I.~M., {Arag{\'o}n-Salamanca}, A., {De Lucia}, G., {et~al.} 2008,
  \mnras, 387, 1253

\bibitem[{{White}(1976)}]{1976MNRAS.174...19W}
{White}, S.~D.~M. 1976, \mnras, 174, 19

\bibitem[{{Yasuda} {et~al.}(2001){Yasuda}, {Fukugita}, {Narayanan}, {Lupton},
  {Strateva}, {Strauss}, {Ivezi{\'c}}, {Kim}, {Hogg}, {Weinberg}, {Shimasaku},
  {Loveday}, {Annis}, {Bahcall}, {Blanton}, {Brinkmann}, {Brunner}, {Connolly},
  {Csabai}, {Doi}, {Hamabe}, {Ichikawa}, {Ichikawa}, {Johnston}, {Knapp},
  {Kunszt}, {Lamb}, {McKay}, {Munn}, {Nichol}, {Okamura}, {Schneider},
  {Szokoly}, {Vogeley}, {Watanabe}, \& {York}}]{2001AJ....122.1104Y}
{Yasuda}, N., {Fukugita}, M., {Narayanan}, V.~K., {et~al.} 2001, \aj, 122, 1104

\bibitem[{{Zhang} {et~al.}(2015){Zhang}, {McKay}, {Bertin}, {Jeltema},
  {Miller}, {Rykoff}, \& {Song}}]{2015PASP..127.1183Z}
{Zhang}, Y., {McKay}, T.~A., {Bertin}, E., {et~al.} 2015, \pasp, 127, 1183

\bibitem[{{Zibetti} {et~al.}(2005){Zibetti}, {White}, {Schneider}, \&
  {Brinkmann}}]{2005MNRAS.358..949Z}
{Zibetti}, S., {White}, S.~D.~M., {Schneider}, D.~P., \& {Brinkmann}, J. 2005,
  \mnras, 358, 949

\end{thebibliography}
